\documentclass[12pt]{article}

\usepackage{amsmath,amssymb,epsfig,euscript,bbold}
\usepackage[matrix,arrow]{xy}
\usepackage{xspace}

\numberwithin{equation}{section}       

\newcommand{\dd}{\mathrm{d}}
\newcommand{\ii}{\mathrm{i}}
\newcommand{\conj}{\mathrm{c}}

\newcommand{\bbZ}{\mathbb{Z}}
\newcommand{\bbR}{\mathbb{R}}
\newcommand{\bbC}{\mathbb{C}}

\newcommand{\id}{\mathbb{1}}

\DeclareMathOperator{\SU}{\mathit{SU}}
\DeclareMathOperator{\SO}{\mathit{SO}}
\DeclareMathOperator{\Symp}{\mathit{Sp}}
\DeclareMathOperator{\Spin}{\mathit{Spin}}

\DeclareMathOperator{\su}{\mathit{su}}
\DeclareMathOperator{\symp}{\mathit{sp}}
\DeclareMathOperator{\sspin}{\mathit{spin}}

\newcommand{\rep}[1]{\mathbf{#1}}

\DeclareMathOperator{\tr}{tr}
\DeclareMathOperator{\Tr}{Tr}

\DeclareMathOperator{\re}{Re}
\DeclareMathOperator{\im}{Im}

\DeclareMathOperator{\vol}{vol}

\newcommand{\smallsec}[1]{\medskip\noindent\underline{\textbf{#1}}:}

\newcommand{\be}{\begin{equation}}
\newcommand{\ee}{\end{equation}}
\newcommand{\bea}{\begin{eqnarray}}
\newcommand{\eea}{\end{eqnarray}}

\newcommand{\nn}{\nonumber \\}
\newcommand{\reef}[1]{(\ref{#1})}

\newcommand{\wed}{\wedge}

\newcommand{\CY}{CY\xspace}
\newcommand{\Ka}{K\"{a}hler\xspace}

\newcommand{\spin}{\ensuremath{\mathit{Spin}(7)}\xspace}

\newcommand{\nij}{Nijenhuis\xspace}
\newcommand{\e}{\epsilon}

\newcommand{\trsp}{{\mathrm T}}

\newcommand{\ex}{{\mathrm e}}
\newcommand{\diff}{{\mathrm d}}

\newcommand{\de}{\partial}

\newcommand{\p}[1]{(\ref{#1})}

\begin{document}


\begin{titlepage}

\vfill

\begin{flushright}
QMUL-PH-03-01\\
hep-th/0302158\\
\end{flushright}

\vfill

\begin{center}

\baselineskip=16pt

{\Large\bf Superstrings with Intrinsic Torsion}

\vskip 1cm 

Jerome P. Gauntlett$^{1}$, Dario Martelli$^{2}$ and Daniel
Waldram$^{3}$ 

\vskip 1cm

\textit{Department of Physics\\ Queen Mary, University of London\\
  Mile End Rd, London E1 4NS, U.K.}  

\end{center}

\vfill

\begin{center}
\textbf{Abstract}
\end{center}

\begin{quote}
We systematically analyse the necessary and sufficient conditions for
the preservation of supersymmetry for bosonic geometries of the 
form $\bbR^{1,9-d}\times M_d$, in the common NS-NS sector of 
type II string theory and also type I/heterotic string theory. 
The results are phrased in terms of the intrinsic torsion of $G$-structures
and provide a comprehensive classification of static supersymmetric
backgrounds in these theories. 
Generalised calibrations naturally appear since the 
geometries always admit NS or type I/heterotic 
fivebranes wrapping calibrated cycles.
Some new solutions are presented. In particular we
find $d=6$ examples with a fibred structure which preserve ${\cal N}=1,2,3$
supersymmetry in type II and include compact type~I/heterotic geometries. 
\end{quote}

\vfill 

\hrule width 5.cm \vskip 5mm 
{\small
\noindent $^1$ E-mail: j.p.gauntlett@qmul.ac.uk \\
\noindent $^2$ E-mail: d.martelli@qmul.ac.uk \\
\noindent $^3$ E-mail: d.j.waldram@qmul.ac.uk}

\end{titlepage}


\section{Introduction}


Supersymmetric backgrounds of string/M-theory 
with non-vanishing fluxes are currently
an active area of study for at least two reasons. Firstly,
they provide a framework for searching for new models with
attractive phenomenology
and secondly, they appear in generalisations of
the AdS/CFT correspondence. For both applications
a detailed mathematical understanding of the kinds of geometry that 
can arise is important for further elucidating physical results. 
Such an understanding can also lead to 
new methods for constructing explicit examples.

Here we will analyse supersymmetric geometries of the
common NS-NS sector of type IIA and IIB supergravity. That is, we
consider non-vanishing dilaton $\Phi$ and three-form $H$ but with all
R-R fields and fermions set to zero. The closely related
type I and heterotic geometries which allow in addition non-trivial
gauge fields will also be considered. Let us introduce the basic
conditions. A type II geometry will preserve supersymmetry if and only if
there is at least one $\epsilon^{+}$ or $\epsilon^-$ satisfying 
\begin{equation}
\label{susy}
\begin{aligned}
   \nabla^\pm_M \epsilon^\pm \equiv \left( \nabla_M \pm 
         \tfrac{1}{8}H_{MNP}\Gamma^{NP} \right)\epsilon^\pm 
      &= 0 , \\
   \left(\Gamma^M\partial_M\Phi \pm 
         \tfrac{1}{12}H_{MNP} \Gamma^{MNP}
         \right) \epsilon^\pm 
      &= 0 ,
\end{aligned}
\end{equation}
where for type IIB (respectively IIA) $\e^\pm$ are two Majorana--Weyl
spinors of $\Spin(1,9)$ of the same (respectively opposite)
chirality and $\nabla$ is the Levi--Civita connection. Geometrically
$\nabla^\pm$ are connections with totally anti-symmetric torsion given
by $\pm\frac12H$. Locally the three-form is given by $H=\dd B$ and
hence satisfies the Bianchi identity
\begin{equation}
\label{bianchi}
   \dd H = 0 .
\end{equation}
For heterotic/type I string theory, the bosonic field content also
includes a gauge field $A$, with field strength $F$, in the adjoint of
$E_8\times E_8$ or $SO(32)/\bbZ_2$. We choose conventions where a
geometry preserves supersymmetry if there is at least one spinor
$\epsilon^+$ satisfying~\eqref{susy} and, in addition, the gaugino
variation vanishes, 
\bea
\label{gaugino}
\Gamma^{MN} F_{MN}\e^+=0 .
\eea
The Bianchi identity reads
\bea
\dd H = 2\alpha'(\Tr F\wedge F - \tr R\wedge R)
\eea
where the second term on the right hand side is the leading string
correction to the supergravity expression. The equations of motion for
these conventions can be found in Appendix~\ref{app:conv}.

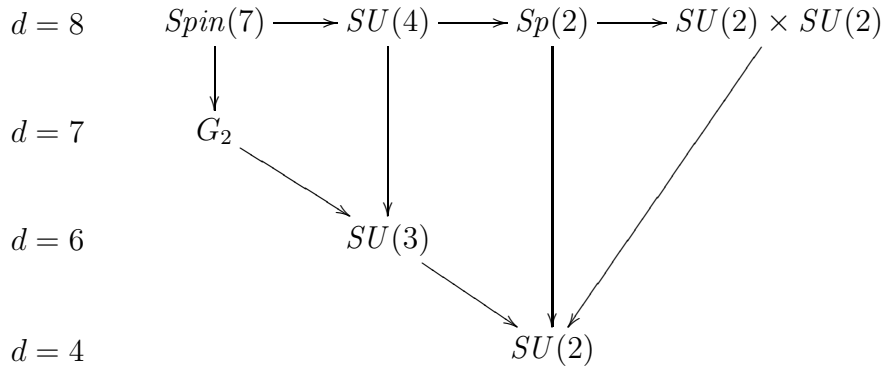
\begin{figure}[!th]
\begin{equation*}
   \xymatrix{
      d=8 & \Spin(7) \ar[r] \ar[d] & \SU(4) \ar[r] \ar[dd] & 
         \Symp(2) \ar[r] \ar[ddd] & \SU(2)\times\SU(2) \ar[dddl] \\
      d=7 & G_2 \ar[dr] & & & \\
      d=6 & & \SU(3) \ar[dr] & & \\
      d=4 & & & \SU(2) 
   } 
\end{equation*}
\caption{Special holonomies of manifolds in  $d$-dimensions
with covariantly constant spinors with respect to either the 
Levi-Civita connection or a connection with totally anti-symmetric 
torsion $H$. Only the minimal ``canonical'' dimension $d$ is presented.
The arrows represent the different ways the groups can 
be embedded in each other.}
\label{fig}
\end{figure}

The geometries we consider here will be of the form
$\bbR^{1,9-d}\times M_d$, and hence with $H, \Phi$ only non-vanishing
on $M_d$. When $d=9$ the analysis covers the most general static
geometries. As is well known, for the special case when
$H=\Phi=0$, the necessary and sufficient conditions for
preservation of supersymmetry is that $M_d$ admits at least one
covariantly constant spinor and hence has special holonomy. Apart from
the trivial  case of flat space this gives rise to the possibilities
presented in figure~\ref{fig}. These manifolds are all Ricci-flat and
hence they automatically also
solve the supergravity equations of motion\footnote{Note that there 
are also higher order corrections to the equations of motion
that give rise to tadpoles for type IIA in $d=8$ and IIB in $d=6$ (via
$F$-theory) \cite{Sethi:1996es}.  The tadpoles can often be
cancelled by the addition of spacetime filling strings or D3-branes,
respectively. Here we shall not explicitly refer to
these corrections further.}. Note that figure~\ref{fig} only presents the minimal
``canonical'' dimension $d$ of the manifold in order for it to have the
corresponding special holonomy. It is also possible to have manifolds of
higher dimension with the same
special holonomy group: when
$H=\Phi=0$, after going to the covering space, the resulting geometries are
simply direct products of special holonomy manifolds in the canonical 
dimensions given in figure~\ref{fig} with one or more flat directions.

The analysis of a set of necessary and sufficient conditions
for the preservation of supersymmetry in certain cases where $H$
and $\Phi$ are non-zero was initiated some time ago in
\cite{strominger} (see also \cite{hull,dewit}). 
In general, from the first condition in~\eqref{susy}, 
it is necessary that there is at least one spinor
which is covariantly constant with respect to one of the connections
$\nabla^\pm$ with totally anti-symmetric torsion, $\nabla^+$ say. 
This is equivalent to requiring that $\nabla^+$ has holonomy given
by one of the groups in
figure~\ref{fig}. As we discuss in more detail below this
implies the existence of various invariant forms on $M_d$ satisfying
certain differential constraints. The second equation
in~\eqref{susy} then imposes additional conditions on the
forms. Finally, one shows that the existence of such a set of forms
with constraints is in fact sufficient for the existence of one or
more solutions to the supersymmetry conditions~\eqref{susy}.  

It is also important to know what extra conditions
are required in order that the geometry solves the equations of motion.
By analysing the integrability conditions of~\p{susy}, 
it was proved in~\cite{Gauntlett:2002sc} for the entire class of 
geometries under consideration, that it is only necessary to impose
the Bianchi identity \p{bianchi}. Note, it was actually shown that one
needs to impose the Bianchi identity for $H$ and the $H$ equation of
motion. However, the expression for $H$ implied by supersymmetry, to
be discussed below and given in \p{exp}, implies that the $H$ equation
of motion is automatically satisfied so only the Bianchi identity is
required. 

Recently it has been appreciated that the necessary and sufficient
conditions derived in \cite{strominger}, which just analysed
the $SU(n)$ cases in $d=2n$, can also be phrased in terms of 
$G$-structures, and this has allowed a number of 
generalisations~\cite{Gauntlett:2001ur,Friedrich:2001nh,
Friedrich:2001yp,Iv3,Gauntlett:2002sc}. 
Similar ideas have been used to analyse other
supergravity solutions in
\cite{the5,dan,jerstas,Kaste:2003dh,Behrndt:2003uq}. 
The invariant forms on $M_d$ define the $G$-structure, while the
differential conditions correspond to restricting the class of the
intrinsic torsion of the $G$-structure. We will briefly review some
aspects of $G$-structures later, but we refer to, e.g.,~\cite{sal} for
further details.  The necessary and sufficient conditions for the
$G_2$ in $d=7$
\cite{Gauntlett:2001ur,Friedrich:2001nh,Friedrich:2001yp} and \spin in
$d=8$ \cite{Iv3} cases have also been analysed from this point of
view. Thus, when only $\nabla^+$, say, has special holonomy $G$, we
now have a fairly complete set of results, assuming that $M_d$ has the
canonical dimension for $G$ as given in figure~\ref{fig}. We shall
review all known cases including the results of
\cite{strominger}. Note the $\SU(3)$ case was also recently reviewed
in detail from the new perspective of intrinsic torsion
in~\cite{dallagata}. One new result of this paper will be to analyse
the remaining two cases in $d=8$ when $\nabla^+$ has holonomy $Sp(2)$
or $SU(2)\times SU(2)$.  

One can also ask what happens when $M_d$ does not have
the canonical dimension for $G$. For example, we might consider
geometries of the form $\bbR^{1,2}\times M_7$, with $M_7$
admitting two Killing spinors leading to $M_7$ having
an $SU(3)$ structure corresponding to $\nabla^+$ having
$SU(3)$ holonomy. In the case that $H=\Phi=0$, as already
noted, this would necessarily imply that $M_7$ is a direct
product of a flat direction with a Calabi-Yau three-fold.
When $H,\Phi\ne 0$, however, we will show that the geometries
can be more general than simply the direct product of a flat direction
with a six manifold $M_6$ with $SU(3)$ structure of the type derived
in  \cite{strominger}.
In particular, the flat direction can be non-trivially fibred
over $M_6$ with the fibration determined by an Abelian 
$SU(3)$ instanton (i.e. a holomorphic gauge field satisfying the
Donaldson--Uhlenbeck--Yau equation). 

More generally, we will determine the most general static
supersymmetric geometries of the form $\bbR^{1}\times M_9$ preserving
any number of Killing spinors $\e^+$. If there is one Killing spinor the
geometry will have a \spin structure but now in $d=9$. Additional
Killing spinors lead to additional \spin structures or equivalently a
$G$-structure where $G$ is the maximal common subgroup of them
embedded in $\SO(9)$. The $G$-structures that arise are still given by
the groups  as in figure~\ref{fig} but now in $d=9$.
We will show that the most general geometries consist of a number of
flat directions non-trivially fibred over manifolds $M_d$ that possess
$G$-structures in the canonical dimension. The fibration is determined
by Abelian generalised $G$-instantons on $M_d$. 

Another purpose of this paper is to present the new and the known
results in a uniform way. In particular, as emphasised in 
\cite{Gauntlett:2001ur}, the
expression for the three-form can always be expressed in terms of the
$G$-structure in a way related to ``generalised
calibrations'' \cite{Gutowski:1999iu,gpt}. 
Specifically we always have an expression of the form 
\begin{equation} 
\label{exp}
   *H = \ex^{2\Phi} \dd\left( \ex^{-2\Phi} \Xi \right)
\end{equation}
where $\Xi$ is an invariant form which specifies, at least partially, 
the $G$-structure. Generalised calibrations extend the original
definition of a calibration form to cases where the background has
non-vanishing fluxes. In particular a generalised calibration form,
here $\Xi$, is no longer closed and its exterior derivative is related
to the flux, here $H$ (and the dilaton $\Phi$)
as in \eqref{exp}. The physical significance of
generalised calibrated cycles is that they minimise the energy
functional of a brane wrapping the cycle in the presence of the
fluxes. 

The reason that~\eqref{exp} might have been anticipated is as follows.  
First one notes that the type of geometries under discussion arise as
solutions describing NS fivebranes wrapping supersymmetric cycles in
manifolds of special holonomy including the full back-reaction of the
brane on the geometry. To see this first recall that the geometry of
an unwrapped NS fivebrane is a product of $\bbR^{1,5}$ along the
world-volume of the fivebrane with a transverse four-dimensional space
with non-vanishing $H$ and $\Phi$. In addition, we know that a {\it probe}
fivebrane with world-volume $\bbR^{1,5-p}\times \Sigma_p$ will be
supersymmetric if $\Sigma_p$ is a calibrated $p$-cycle in some special
holonomy background. When we go beyond the probe approximation and
consider the back reaction of the fivebrane on the geometry, we thus expect a
geometry of the form $\bbR^{1,5-p}\times M_{p+4}$ with non-vanishing
$H$ and $\Phi$. This is precisely the type of geometry we are
considering. Now, on physical grounds, we know that we can always add
a second probe brane without breaking supersymmetry provided it is
wrapping a cycle calibrated by the same calibration form $\Xi$ as the
original probe brane. This implies that as we switch on the back
reaction, $\Xi$ should still be a calibrating form, though now, since
$H$ and $\Phi$ are non-zero, it is a generalised calibration. In other
words, if the original probe brane wraps a cycle calibrated by a
calibration form $\Xi$, the final geometry $M_{p+4}$ should admit the
corresponding generalised calibration form, that is $\Xi$
satisfying~\eqref{exp}.  

In table~\ref{green} we have listed for each of the different holonomies
of $\nabla^+$, in the canonical dimensions,
the corresponding type of calibrated cycle
that a NS-fivebrane wraps in order to give the geometry. We have also
included the number of minimal $Spin(d)$ spinors preserved in 
each case. Note that for the $d=4$ and $d=8$ cases we have listed the 
six- and two-dimensional chirality of the preserved
supersymmetry. Also $CY_n$ corresponds to a Calabi-Yau $n$-fold and
$HK_2$ to a hyper-\Ka manifold in $d=8$.
\begin{table}[!th]
\begin{center}
\setlength{\tabcolsep}{0.45em}
\begin{tabular}{|c|c|cc|c|c|c|}
\hline
dim($M$) & ${\cal N}$ & 
   Hol($\nabla^{+}$) & Hol($\nabla^{-}$)& $G$-structure & 
   calibrated cycle \\ 
\hline\hline
4 & (1,0)     & $\SU(2)$    & $\Spin(4)$   & $\SU(2)$   & 
   point in $CY_2$ \\
6 & 1         & $\SU(3)$    & $\Spin(6)$   & $\SU(3)$   & 
   \Ka-2 in $CY_3$ \\
7 & 1        & $G_2$       & $\Spin(7)$   & $G_2$      & 
   associative  in $G_2$ \\
8 & (4,0)  & $\SU(2)^2$  & $\Spin(8)$ & $\SU(2)^2$ & 
   $CY_2$ and/or $CY_2'$ in $CY_2\times CY_2'$ \\
8 & (3,0)  & $\Symp(2)$  & $\Spin(8)$   & $\Symp(2)$ & 
   Quaternionic  in $HK_2$ \\
8 & (2,0)  & $\SU(4)$    & $\Spin(8)$   & $\SU(4)$   & 
   \Ka-4  in $CY_4$ \\
8 & (1,0)  & $\Spin(7)$  & $\Spin(8)$   & $\Spin(7)$ & 
   Cayley in $\Spin(7)$  \\
\hline
\end{tabular}
\end{center}
\caption{$G$-structures when $\nabla^+$ has special holonomy.}
\label{green}
\end{table}

It is interesting to note that the more general geometries in $d=9$
mentioned above, with a number of flat directions fibred
over $M_d$, have a fascinating interpretation in this regard.
In particular, the flat directions correspond to directions along 
the world-volume of the fivebrane wrapping a flat direction, 
and so it is surprising that supersymmetry does not require the 
fibration to be trivial. Note that this interpretation is mirrored in
the refined version of~\p{exp} for the flux that one obtains in $d=9$: 
\begin{equation} 
\label{exp9}
   *H = \ex^{2\Phi} 
   \dd\left( \ex^{-2\Phi} \Xi\wedge K^1\wedge\dots\wedge K^{9-d} \right)
\end{equation}
where $\Xi$, $K^i$ (partly) determine the $G$-structure, with $K^i$
one forms corresponding to the flat directions of the fivebrane.

The fact that the geometries all satisfy calibration conditions of the 
form~\eqref{exp} connects with a simple vanishing
theorem for compact backgrounds~\cite{FGW,dWSHD}. Consider the dilaton
equation of motion~\eqref{Phieom} as given in Appendix~\ref{app:conv}
for the type I case, setting $F=0$ for the type II case. Suppose $M_d$
is compact, integrating the equation of motion gives  
\begin{equation}
\label{nogo1}
   \int_{M_d} \ex^{-2\Phi} H \wedge *H
      + 2\alpha' \int_{M_d} \ex^{-2\Phi} \Tr F \wedge *F 
      = 0 .
\end{equation}
Since the integrand in each term is positive semi-definite, we must
have $H=F=0$ and hence $\Phi$ is constant. Thus, we see that there are
no compact solutions in type II and type I supergravities with
non-zero flux $H$ and dilaton. This vanishing theorem can of course be
evaded if one includes leading-order heterotic/type~I string
corrections which introduce additional $\tr R^2$ terms in the dilaton
equation of motion. 

The theorem is reproduced in the special supersymmetric
sub-case as a consequence of the calibration condition~\eqref{exp} and the 
Bianchi identity. This is a reflection of the general
result~\cite{dWSHD,Gauntlett:2002sc} that the equations of motion are
implied by the preservation of supersymmetry and the Bianchi
identity. One has 
\begin{equation}
\label{nogo2}
   \int_{M_d} \ex^{-2\Phi}H\wedge *H 
      = \int_{M_d} H \wedge \dd(\ex^{-2\Phi} \Xi) 
      = -  \int_{M_d} \ex^{-2\Phi} \dd H \wedge \Xi . 
\end{equation}
The simplest case~\cite{Gauntlett:2002sc} is when $\dd H=0$ (as is
true for any type~II background). We then have $H=\Phi=0$ by the same
positivity argument as above. (This simplifies and
extends\footnote{Note that~\cite{ivpapone} includes results for the
  $SU(n)$ case when $\dd H\ne 0$.} an earlier vanishing theorem that
was given for the $\SU(n)$ cases only in \cite{ivpapone}). In the case
of Type~I supergravity, one finds that the Bianchi identity together
with the conditions on $F$ for supersymmetry
(see~\eqref{eq:instanton} below) imply the last expression
in~\eqref{nogo2} can be rewritten as minus the second term 
in~\eqref{nogo1}, and again we find $H=\Phi=F=0$.

Until this point the discussion has focused on geometries admitting
one or more Killing spinors of the same type, $\e^+$, say.
This covers all static cases of the type I/heterotic theories.
However, for the type II theories when $H$ and $\Phi$ are non-zero,
there are solutions to~\eqref{susy} for both $\epsilon^+$ and
$\epsilon^-$, if both connections $\nabla^+$ and $\nabla^-$
have special holonomy. This means that the general classification of
supersymmetric geometries indicated in table~\ref{green}, as well
as the generalisations to $d=9$, can be refined.
In \cite{Gauntlett:2001ur} we analysed the different ways in
which probe fivebranes can wrap calibrated cycles in manifolds of
special holonomy and determined the holonomies of 
$\nabla^\pm$ that are expected in the corresponding supergravity solutions,
after including the back reaction. The results are summarised in 
table~\ref{blue}. In these cases, $\e^\pm$ each define a 
different structure with groups $G^\pm$. Equivalently, together they define
a single structure with group $G$ which is the maximal common 
subgroup of the two embedded in $\SO(d)$, and this is also listed in
table~\ref{blue}. 
{}It is noteworthy that from the 
wrapped fivebrane perspective, in all cases
this minimal $G$-structure is the same as the 
holonomy of the initial special holonomy manifold that one started
with. Since both $\e^\pm$ are required to define the $G$-structure,
unlike the $G^\pm$-structures, it is not covariantly constant 
with respect to a connection with totally anti-symmetric torsion.

\begin{table}[!th]
\begin{center}
\setlength{\tabcolsep}{0.45em}
\begin{tabular}{|c|c|c|cc|c|c|}
\hline
dim($M$) & ${\cal N}_{\mathrm{IIB}}$ & ${\cal N}_{\mathrm{IIA}}$ & 
   Hol($\nabla^{+}$) & Hol($\nabla^{-}$)& $G$-structure & 
   calibrated cycle \\ 
\hline\hline
4 & (1,1) & (2,0) & $\SU(2)$   & $\SU(2)$   & $\{\id\}$  & 
   point in $\bbR^4$         \\
6 & 2     & 2     & $\SU(3)$   & $\SU(3)$   & $\SU(2)$   & 
   \Ka-2  in $CY_2$        \\
7 & 2     & 2     & $G_2$      & $G_2$      & $\SU(3)$   & 
   SLAG-3 in $CY_3$       \\
8 & (2,2) & (4,0) & $\SU(4)$   & $\SU(4)$   & $\SU(3)$   & 
   \Ka-4  in $CY_3$       \\
8 & (4,0) & (2,2) & $\SU(4)$   & $\SU(4)$   & $\SU(2)^2$ & 
   \Ka-2$\times$\Ka-2 in $CY_2\times CY_2'$         \\
8 & (3,0) & (2,1) & $\SU(4)$   & $\Spin(7)$ & $\Symp(2)$ & 
   C-LAG-4 in $HK_2$      \\
8 & (2,0) & (1,1) & $\Spin(7)$ & $\Spin(7)$ & $\SU(4)$   & 
   SLAG-4  in $CY_4$      \\
8 & (1,1) & (2,0) & $\Spin(7)$ & $\Spin(7)$ & $G_2$      & 
   coassociative in $G_2$\\
\hline
\end{tabular}
\end{center}
\caption{$G$-structures in type II theories when both $\nabla^\pm$ have
special holonomy.}
\label{blue}
\end{table}

The particular class of geometries with
$\nabla^\pm$ each having $G_2$ holonomy, with a common $SU(3)$ subgroup was
analysed in detail in \cite{Gauntlett:2002sc}.
The necessary and sufficient conditions on the $\SU(3)$ structure
in order that the geometry preserves supersymmetry were presented. 
This case is associated with fivebranes 
wrapping SLAG three-cycles in manifolds with $\SU(3)$ holonomy. 
It was also shown that the three-form flux can be expressed as a generalised 
calibration associated with a (3,0) form, as expected for a special
Lagrangian cycle. This result again refines that of \p{exp} in a way
expected from physical considerations.  
Here we shall extend the analysis of \cite{Gauntlett:2002sc} 
to cover all cases discussed in \cite{Gauntlett:2001ur}.

Table~\ref{blue} lists the geometries associated with
fivebranes wrapping calibrated cycles.  Note that explicit 
solutions corresponding to three more cases were 
discussed in \cite{george}: $\nabla^+$ has $\Symp(2)$
holonomy, while $\nabla^-$ has $\Spin(7)$, $\SU(4)$ or $\Symp(2)$ holonomy.
They correspond to fivebranes wrapping certain quaternionic planes in $\bbR^8$.
Such calibrations are linear and it is plausible that the solutions found
in \cite{george} are the most general kinds of solution. 
In any case, we will not consider these cases further in this paper.

The geometries listed in table~\ref{blue} are all
in their ``canonical'' dimension.
We will argue that they can be generalised to $d=9$, as before,
by adding a number of flat directions.
In order that both $\e^+$ and $\e^-$ Killing spinors survive, the
fibration must be given by a generalised instanton with respect to the
common $G$-structure. 

It is natural to wonder if supersymmetric geometries admitting
both $\e^+$ and $\e^-$ Killing spinors are necessarily of the type 
given in table~\ref{blue}. We shall present an interesting explicit
example in $d=6$ which shows that this is not the case. The example is
a torus $T^2$ non-trivially fibred over a flat $\bbR^4$ base with
non-vanishing dilaton. For a particular carefully chosen fibration we
show that $\nabla^+$ has $SU(3)$ holonomy while $\nabla^-$ has $SU(2)$
holonomy. This solution thus preserves twelve supercharges which
corresponds to ${\cal N}=3$ supersymmetry in the remaining  four spacetime
dimensions.  It would be interesting to see how it is related to the
type IIB solutions preserving the same amount of supersymmetry with
both R-R and NS-NS fluxes presented in \cite{Frey:2002hf}. 

In this paper we will not explicitly present many detailed proofs
since the arguments follow the same lines as those
in \cite{Gauntlett:2001ur,Gauntlett:2002sc}, and also because
we do not want to obscure the main results. 
The plan of the rest of the paper
is as follows. In section~\ref{sec:Gstructure} we review
$G$-structures and their intrinsic torsion. In
section~\ref{sec:canon} we discuss the geometries summarised in
table~\ref{green}. We also comment on the additional constraints
arising in type I/heterotic string theory. Section~\ref{sec:gen}
analyses the general supersymmetric geometries in $d=9$ when one of
the connections $\nabla^\pm$ has special holonomy,  which generalises
the geometries of table~\ref{green}. In section~\ref{sec:ex1} we
present some simple explicit  solutions of the type discussed in
section~\ref{sec:gen} including candidate heterotic/type~I
compactifications based on fibrations over $K3$ surfaces that preserve
eight supersymmetries. Section~\ref{sec:+-} discusses the 
cases summarised in table~\ref{blue} when both $\nabla^+$ and $\nabla^-$
have special holonomy. Section~\ref{sec:ex2} presents some further
explicit solutions in $d=6$ including a type~II example preserving
12 supersymmetries corresponding to ${\cal N}=3$ supersymmetry 
and candidate heterotic/type~I compactifications based on fibrations
over $K3$ surfaces that preserve four
supersymmetries. Section~\ref{sec:concl} concludes with some
discussion and a summary of our main results.


\section{$G$-structures in canonical dimension} 
\label{sec:Gstructure}


It will be useful first to recall some aspects of the classification
of $G$-structures (for further details see e.g. \cite{sal}). 
A manifold $M_d$ admits a $G$-structure if its
frame bundle admits a sub-bundle with fibre group $G$. This implies
that all tensors and, when appropriate, spinors on $M_d$ can be decomposed
globally into representations of $G$. A $G$-structure is typically
equivalent to the existence of a set of globally defined
$G$-invariant tensors, or alternatively a set of globally defined
$G$-invariant spinors. In particular, when $G\subset\Spin(d)$ as is the
case for $G$-invariant spinors, the structure defines a metric, since
the corresponding sub-bundle of the frame bundle can be viewed as a
set of orthonormal frames. 

The $G$-structure is classified by the
intrinsic torsion. When $G\subset\Spin(d)$ this is a measure of the
failure of the tensors/spinors to be covariantly
constant with respect to the Levi--Civita connection of the metric defined
by the structure. As a result, all of the components of the
intrinsic torsion are encoded in derivatives of the invariant
tensors/spinors. Furthermore, the intrinsic torsion, $T$, then takes values in
$\Lambda^1\otimes g^\perp$ where $\Lambda^p$ is the space of $p$-forms
and $g^\perp\oplus g=\sspin(d)$ where $g$ is the Lie algebra of $G$. The intrinsic
torsion can then be decomposed into irreducible $G$-modules, 
$T\in\oplus_i {\cal W}_i$. We will denote specific components of $T$ in 
each module ${\cal W}_i$ by $W_i$. Only if the intrinsic torsion
completely vanishes does the manifold have $G$-holonomy.    

For a supersymmetric background $(M_d,g_d,H,\Phi)$, where $g_d$ is the
metric on $M_d$, we need some non-trivial globally 
defined spinors satisfying~\eqref{susy}. Note, the spinors are
globally defined since $\nabla^\pm \e^\pm=0$ implies they have
constant norm, which we take to be unity, $\bar\e^\pm \e^\pm=1$, and
so are nowhere vanishing. This necessarily defines a $G$-structure with
$G\subset\Spin(d)$. The possible groups $G$ are precisely the possible
special holonomy groups appearing in figure~\ref{fig}. The necessary
and sufficient conditions for solutions of the particular
supersymmetry constraints~\eqref{susy} then translate into the
$G$-structure being of a particular type with certain components of
the intrinsic torsion vanishing. Since $G\subset\Spin(d)$ the metric
$g_d$ is completely determined by the $G$-structure. Similarly, one
finds expressions for $H$ and $\Phi$ in terms of the intrinsic
torsion of the $G$-structure.

In this section we will summarise the definition of the structures and
how the generic intrinsic torsion is encoded in each case. We will only
consider the structures in their canonical dimensions: \spin in $d=8$,
$G_2$ in $d=7$, etc. It is straightforward to generalise to the case that
the structure is in a higher dimension (for an example, see appendix E of
\cite{jerstas}).
In the following sections we then turn to the particular necessary and
sufficient conditions on the structure for supersymmetry. 

\smallsec{$\SU(n)$-structure in $d=2n$} 
The structure is completely specified by a real two-form $J$ of
maximal rank and a complex $n$-form $\Omega$ satisfying  
\begin{equation}
\label{SUn}
\begin{aligned}
   J \wedge \Omega &= 0 , \\
   \Omega \wedge \bar{\Omega} &= \ii^{n(n+2)}\frac{2^n}{n!}J^n ,
\end{aligned}
\end{equation}
where $J^n$ is defined using the wedge product. Together these define
a metric $g_d$ and an orientation chosen as $\vol=J^n/n!$.
Raising an index on $J$ using this metric defines an almost complex
structure satisfying $J^2=-\id$. With respect to this 
almost complex structure, $\Omega$ is an $(n,0)$-form while the
two-form $J$ is of type $(1,1)$. Furthermore the metric $g_d$ is
almost Hermitian. Note that the almost complex structure is actually
determined solely by the choice of $\Omega$ and is independent of the
two-form~$J$. 

{}For generic $\SU(n)$ structures, the intrinsic torsion decomposes into five
modules $\mathcal{W}_i$~\cite{sal,grayhervella,chiossi}. Consider for
instance $\SU(4)$. The adjoint representation of $Spin(8)$ decomposes as
$\rep{28}\to\rep{1}+\rep{6}+\rep{\bar{6}}+\rep{15}$ where $\rep{15}$ is
the adjoint representation of $\SU(4)$, and so the remaining
representations correspond to $\su(4)^\perp$. The one-form $\Lambda^1$
representation decomposes as $\rep{8}\to\rep{4}+\rep{\bar 4}$. We then have 
\begin{equation}
   T \in \Lambda^1 \otimes \su(n)^\perp = 
      \mathcal{W}_1 \oplus \mathcal{W}_2 \oplus \mathcal{W}_3 
         \oplus \mathcal{W}_4 \oplus \mathcal{W}_5 .
\end{equation}
where the corresponding $\SU(4)$ representations of $\mathcal{W}_i$
are given by
\begin{equation}
   (\rep{4}+\rep{\bar{4}}) \times (\rep{1}+\rep{6}+\rep{\bar{6}})
       = (\rep{4}+\rep{\bar{4}}) + (\rep{20}+\rep{\bar{20}})
          + (\rep{20}+\rep{\bar{20}}) + (\rep{4}+\rep{\bar{4}})
          + (\rep{4}+\rep{\bar{4}}) .
\end{equation}
For $n=2$ and $n=3$ the corresponding representations are 
\begin{equation}
\begin{aligned}
   (\rep{2}+\rep{\bar{2}}) &\times (\rep{1}+\rep{1}+\rep{1})
       = (\rep{2}+\rep{\bar{2}}) + (\rep{2}+\rep{\bar{2}})
          + (\rep{2}+\rep{\bar{2}}) &&, \\
   (\rep{3}+\rep{\bar{3}}) &\times (\rep{1}+\rep{3}+\rep{\bar{3}}) 
       = (\rep{1}+\rep{1}) + (\rep{8}+\rep{8})
          + (\rep{6}+\rep{\bar{6}}) + (\rep{3}+\rep{\bar{3}})
          + (\rep{3}+\rep{\bar{3}}) &&,  
\end{aligned}
\end{equation}
respectively. In particular, for $n=2$ the modules $\mathcal{W}_1$ and
$\mathcal{W}_3$ are absent. For $n=3$ note that the $\mathcal{W}_1$
and $\mathcal{W}_2$ modules can be further decomposed into real
modules $\mathcal{W}_1^\pm$ and $\mathcal{W}_2^\pm$ as discussed in
detail in \cite{chiossi}.  

Each component of the intrinsic torsion $W_i\in\mathcal{W}_i$ can be
given in terms of the exterior derivative of $J$ or $\Omega$, or in one
case both. Generically, we have the decompositions
\begin{equation}
\begin{aligned}
   \dd J &\in \mathcal{W}_1 \oplus \mathcal{W}_3 
      \oplus \mathcal{W}_4 , \\
   \dd \Omega &\in \mathcal{W}_1 \oplus \mathcal{W}_2 
      \oplus \mathcal{W}_5 .
\end{aligned}
\end{equation}
Explicitly, since $J$ is a $(1,1)$-form, $\dd J$ has a $(3,0)$ piece
and a $(2,1)$ piece (plus the complex conjugates). The former 
defines an irrep of $SU(n)$ and gives the $W_1$ component of $T$. The
latter splits into a primitive $\dd J^{(2,1)}_0$-form, i.e.
 one satisfying $J\lrcorner\, \dd J^{(2,1)}_0=0$, giving $W_3$, plus a
$(1,0)$-form, giving $W_4$, and which can be written as
\begin{equation}
   W_4 \equiv J \lrcorner\, \dd J .
\end{equation}
The same expression appears in characterising any almost Hermitian
metric and is known as the Lee form (of $J$). Here we have introduced
the notation $\omega\lrcorner\,\nu$ which contracts a $p$-form
$\omega$ into a $(n+p)$-form $\nu$ via
\begin{equation}
   (\omega \lrcorner\,\nu)_{i_1\dots i_n}
       = \frac{1}{p!}\omega^{j_1\dots j_p}
          \nu_{j_1\dots j_pi_1\dots i_n} .
\end{equation}
Similarly, since
$\Omega$ is a $(n,0)$-form, $\dd\Omega$ has a $(n,1)$
piece plus a $(n-1,2)$ piece. Let us first consider $n\ne2$.
Again the former defines an irrep, which
gives $W_5$ and can be written as a Lee form for either $\re\Omega$ or
equivalently $\im\Omega$ 
\begin{equation}
\label{leep}
\begin{aligned}
   W_5 &\equiv \frac{1}{4}(\Omega\lrcorner\,\dd\bar\Omega 
           + \bar\Omega\lrcorner\,\dd\Omega ) , \\
       &= \re\Omega \lrcorner\,\dd(\re\Omega)
      = \im\Omega \lrcorner\,\dd(\im\Omega), \qquad n\ne 2.
\end{aligned}
\end{equation}
The second line is obtained by noting that $\Omega\lrcorner\, \dd\Omega=0$.
In general, the $(n-1,2)$ piece of $\dd\Omega$ 
splits into a primitive piece $\dd\Omega^{(n-1,2)}_0$
giving $W_2$ plus another piece that encodes the same $W_1$ component
of $T$ as $\dd J^{(3,0)}$ due to the second compatibility condition
in~\eqref{SUn}. Note that for $SU(3)$, $W_{1,2}^\pm$ can be defined as
the real and imaginary parts of $W_{1,2}$, respectively. For $SU(2)$,
as noted, the classes ${\cal W}_1$ and ${\cal W}_3$ are absent. In this
case
$W_5$ is still given by the first line of \p{leep}, while $W_2$ is
defined by
\bea
   W_2=\frac{1}{4}(\Omega\lrcorner\,\dd\Omega
      + \bar\Omega\lrcorner\,\dd\bar\Omega) .
\eea

Recall that we have $\SU(n)$-holonomy if all the components of the
intrinsic torsion vanish. In this case the manifold is
Calabi--Yau. Clearly this occurs if and only if $\dd J=\dd\Omega=0$. It
will be useful to note some two further cases. First, the almost
complex structure is integrable if and only if $W_1=W_2=0$. 
Secondly, we note that under a conformal transformation of
the $\SU(n)$-structure, such that $J\to \ex^{2f}J$ and $\Omega \to
\ex^{nf}\Omega$, which implies the metric scales as $g\to \ex^{2f}g$, 
$W_1, W_2$ and $W_3$ are invariant as is the
following combination
\begin{equation}
\label{confinv}
   (2n-2)W_5+(-1)^{n+1} 2^{n-2} n W_4 .
\end{equation}
If this combination together with $W_1$, $W_2$ and $W_3$ all vanish and
$W_{4,5}$ are exact, the manifold is conformally Calabi-Yau. 

\smallsec{$\Spin(7)$-structures in $d=8$} 
The structure is specified by a $\Spin(7)$-invariant Cayley four-form,
$\Psi$, which at any given point in $M_8$ can be written as 
\begin{equation}
\label{Psidef}
\begin{split}
   \Psi &= e^{1234}+e^{1256}+e^{1278}+e^{3456}+e^{3478}+e^{5678} \\
      &\qquad + e^{1357}-e^{1368}-e^{1458}-e^{1467}-e^{2358}
          -e^{2367}-e^{2457}+e^{2468} ,
\end{split}
\end{equation}
where $e^m$ define a local frame and $e^{mnpq}=e^m\wedge e^n\wedge
e^p\wedge e^q$. The structure defines a metric
$g_8=(e^1)^2+\dots+(e^8)^2$ and an orientation which we take to be
$\vol=e^1\wedge\dots\wedge e^8$ implying~$*\Psi=\Psi$. 

The adjoint representation of $\SO(8)$ decomposes under $\Spin(7)$ as
$\rep{28}\to\rep{7}+\rep{21}$, where $\rep{21}$ is the adjoint
representation of $\Spin(7)$. One then finds that the intrinsic
torsion decomposes into two modules~\cite{spin7s} 
\begin{equation}
\begin{aligned}
   T \in \Lambda^1 \otimes \spin^\perp 
      &= \mathcal{W}_1 \oplus \mathcal{W}_2 , \\
   \rep{8} \times \rep{7} &= \rep{8} + \rep{48} . 
\end{aligned}
\end{equation}
The components $W_i$ of $T$ in $\mathcal{W}_i$ are given in terms of
the exterior derivative $\dd\Psi$ as, again decomposing into
$\Spin(7)$ representations,  
\begin{equation}
\begin{aligned}
   \dd \Psi \in \Lambda^5 &\cong \mathcal{W}_1 \oplus \mathcal{W}_2 , \\
   \rep{56} &\to \rep{8} + \rep{48} .
\end{aligned}
\end{equation}
In particular the $W_1$ component in the $\rep{8}$ representation is
given by 
\begin{equation}
   W_1 \equiv \Psi \lrcorner\, \dd\Psi ,
\end{equation}
and is the Lee form for $\Psi$. The $W_2$ component in the
$\rep{48}$ representation is then given by the remaining pieces of
$\dd\Psi$. Note that the \spin manifold has \spin holonomy
only when the intrinsic torsion vanishes which is equivalent to
$\dd\Psi=0$. In addition, under a conformal transformation we have $\Psi \to
\ex^{4f}\Psi$ for some function $f$, which implies that the metric
scales as $g\to \ex^{2f} g$. Such a transformation leaves the $W_2$
component of $T$ invariant while the Lee-form $W_1$ transforms as $W_1 \to
W_1+28\dd f$.   

Given the definition~\eqref{Psidef} one has a number of standard
identities, which will be useful in what follows. We have
\begin{equation}
\begin{aligned}
   \Psi^{m_1m_2m_3p}\Psi_{n_1n_2n_3p}
      &= 6\delta^{m_1m_2m_3}_{n_1n_2n_3} +9 \Psi^{[m_1m_2}{}_{[n_1n_2}
      \delta^{m_3]}_{n_3]} , \\
   \Psi^{m_1m_2p_1p_2}\Psi_{n_1n_2p_1p_2}
      &= 12\delta^{m_1m_2}_{n_1n_2} +4 \Psi^{m_1m_2}{}_{n_1n_2} , \\
   \Psi^{mp_1p_2p_3}\Psi_{np_1p_2p_3}
      &= 42\delta^{m}_{n} . 
\end{aligned}
\end{equation}

\smallsec{$G_2$-structures in $d=7$} 
The structure is specified by an associative three-form $\phi$. In a
local frame this can be given by 
\begin{equation}
\label{g2form}
   \phi = e^{246}-e^{235} - e^{145} - e^{136} 
      + e^{127} + e^{347} + e^{567} .
\end{equation}
This defines a metric $g_7=(e^1)^2+\dots+(e^7)^2$ and an orientation
$\vol=e^1\wedge\dots\wedge e^7$. Explicitly we then have 
\begin{equation}
   *\phi = e^{1234}+e^{1256}+e^{3456}+ e^{1357}-e^{1467}-e^{2367}-e^{2457}.
\end{equation}

The adjoint representation of $\SO(7)$ decomposes as
$\rep{21}\to\rep{7}+\rep{14}$ where $\rep{14}$ is the adjoint
representation of $G_2$. The intrinsic torsion then decomposes into
four modules~\cite{fernandez},  
\begin{equation}
\begin{aligned}
   T \in \Lambda^1 \otimes g_2^\perp 
      &= \mathcal{W}_1 \oplus \mathcal{W}_2 \oplus \mathcal{W}_3
         \oplus \mathcal{W}_4 , \\
   \rep{7} \times \rep{7} 
      &= \rep{1}  + \rep{14} + \rep{27}+ \rep{7} .
\end{aligned}
\end{equation}
The components of $T$ in each module $\mathcal{W}_i$ are encoded in
terms of $\dd\phi$ and $\dd*\phi$ which decompose as 
\begin{equation}
\begin{aligned}
   \dd \phi \in \Lambda^4 
      &\cong \mathcal{W}_1 \oplus \mathcal{W}_3 \oplus \mathcal{W}_4 , \\
   \rep{35} &\to \rep{1}+ \rep{27} + \rep{7}  , \\
   \dd * \phi \in \Lambda^5 
      &\cong \mathcal{W}_2 \oplus \mathcal{W}_4 , \\
   \rep{21} &\to \rep{14} + \rep{7} .
\end{aligned}
\end{equation}
Note that the $W_4$ component in the $\rep{7}$ representation appears
in both $\dd\phi$ and $\dd*\phi$. It is the Lee form, given by
\begin{equation}
   W_4 \equiv \phi \lrcorner\, \dd\phi = -*\phi\lrcorner\, \dd*\phi.
\end{equation}
The $\mathcal{W}_1$ component in the singlet representation can be
written as 
\begin{equation}
   W_1 \equiv *(\phi \wedge \dd\phi) .
\end{equation}
The remaining components of $\dd\phi$ and $\dd*\phi$ encode $W_3$ and
$W_2$ respectively. The $G_2$ manifold has $G_2$ holonomy if and only if
the intrinsic torsion vanishes which is equivalent to $\dd\phi=\dd*\phi=0$. 
Note that under a conformal transformation $\phi \to \ex^{3f}\phi$ the
metric transforms as $g\to \ex^{2f} g$ and hence $*\phi \to
\ex^{4f}*\phi$. Under this transformation $W_1$, $W_2$ and $W_3$ are
invariant, while the Lee-form transforms as $W_4 \to W_4-12 \dd f$. 
Finally, note that $G_2$-structures of the type ${\cal W}_1\oplus{\cal
  W}_3\oplus{\cal W}_4$ are called integrable as one can introduce a $G_2$
Dolbeault cohomology \cite{fernug}.

Again there are a number of useful identities given the
definition~\eqref{g2form}. We have
\begin{equation}
\begin{aligned}
   {*\phi}^{m_1m_2m_3p} {*\phi}_{n_1n_2n_3p}
      &= 6\delta^{m_1m_2m_3}_{n_1n_2n_3} 
        + 9 {*\phi}^{[m_1m_2}{}_{[n_1n_2}\delta^{m_3]}_{n_3]}
        - \phi^{m_1m_2m_3}\phi_{n_1n_2n_3} , \\
   {*\phi}^{m_1m_2p_1p_2} {*\phi}_{n_1n_2p_1p_2}
      &= 8\delta^{m_1m_2}_{n_1n_2} 
         + 2 {*\phi}^{m_1m_2}{}_{n_1n_2} , \\
   {*\phi}^{mp_1p_2p_3} {*\phi}_{np_1p_2p_3} &= 24\delta^{m}_{n},
\end{aligned}
\end{equation}
while
\begin{equation}
\begin{aligned}
   \phi^{m_1m_2p}\phi_{n_1n_2p} 
      &= 2\delta^{m_1m_2}_{n_1n_2} + {*\phi}^{m_1m_2}{}_{n_1n_2} , \\
   \phi^{mp_1p_2}\phi_{np_1p_2} &= 6\delta^{m}_{n} , 
\end{aligned}
\end{equation}
and
\begin{equation}
\begin{aligned}
   \phi^{m_1m_2p} {*\phi}_{n_1n_2n_3p}
      &= \phi^{[m_1}{}_{[n_1n_2}\delta^{m_2]}_{n_3]} , \\
   \phi^{mp_1p_2} {*\phi}_{n_1n_2p_1p_2} &= 4\phi^{m}{}_{n_1n_2} .
\end{aligned}
\end{equation}

\smallsec{$\Symp(n)$-structures in $d=4n$}
The structure is specified by three almost complex structures $J^A$
with $A=1,2,3$ satisfying the algebra
\begin{equation}
   J^{A}\cdot J^{B}=-\delta^{AB}\id +\epsilon^{ABC}J^{C} .
\label{Spstruc}
\end{equation}
Together these define a metric $g_d$. Lowering one index with this metric on
each almost complex structure gives a set of maximal rank two-forms
$J^A$. Note that the $\Symp(n)$-structure could be equally well defined
in terms of these forms. We also have a natural orientation given by
$\vol=(J^A)^{2n}/(2n)!$ for any $J^A$. 

For $n=1$ recall that $\Symp(1)\cong\SU(2)$ and this case has already
been considered above. We can make the correspondence by identifying
$J\equiv J^3$ and $\Omega\equiv J^2+\ii J^1$. In more detail, first
note that one can define nine Lee-forms $L^{AB}\equiv J^A\lrcorner\,
\dd J^B$, but for $\SU(2)$ only the diagonal Lee-forms are
independent, since $J^A\cdot L^{AB}$ is independent of $A$ for each
$B$. The three classes of intrinsic torsion defined above from the
$SU(2)$ point of view, are given by $W_2=\frac{1}{2}(L^{22}-L^{11})$,
$W_4=L^{33}$ and $W_5=\frac{1}{2}(L^{11}+L^{22})$. Note that the
almost complex structure $J^3$ is integrable if and only if
$L^{11}-L^{22}=0$ and similarly for $J^1$ and $J^2$ \cite{gt}. 

The only other case of interest in the context of this paper
is $\Symp(2)$. The adjoint representation of
$SO(8)$ decomposes under $\Symp(2)$ as
$\rep{28}\to3(\rep{1})+3(\rep{5})+\rep{10}$, where $\rep{10}$ is the
adjoint representation of $\Symp(2)$. One then finds that the
intrinsic torsion decomposes into 9 different $\Symp(2)$ modules 
\begin{equation}\label{sp2decomp}
\begin{aligned}
   T \in \Lambda^1 \otimes \symp(2)^\perp
      &= \bigoplus_{i=1}^{9} \mathcal{W}_i , \\
   (\rep{4}+\rep{4})\times(3(\rep{1})+ 3(\rep{5}))
      &= 6(\rep{4}+\rep{4})+3(\rep{16}+\rep{16}) ,
\end{aligned}
\end{equation}
where the notation takes into account that while the torsion is real,
the representations
$\rep{4}$ and $\rep{16}$ are pseudo-real. One can show that all 
the components of $T$ in $\mathcal{W}_i$ are
specified in terms of the exterior derivatives $\dd J^A$. 
Thus the $\Symp(2)$ manifold has $\Symp(2)$ holonomy if and
only if $\dd J^A=0$. In this case six of the nine Lee forms
$L^{AB}\equiv J^A\lrcorner\,\dd J^B$ are linearly independent (this is
actually true for any $\Symp(n)$-structure), and these precisely
correspond to the six $(\rep{4}+\rep{4})$  representations appearing
in \p{sp2decomp}. To be more precise, one can show that
\begin{equation}
\begin{aligned}
   L^{12}+L^{21} &= J^3\cdot (L^{11}-L^{22}) , \\
   L^{31}+L^{13} &= J^2\cdot(L^{33}-L^{11}) , \\
   L^{23}+L^{32} &= J^1\cdot(L^{22}-L^{33}) ,
\end{aligned}
\end{equation}
and hence six independent Lee-forms are given by $L^{11}$, $L^{22}$
and $L^{33}$ and $L^{12}-L^{21}$, $L^{31}-L^{13}$ and
$L^{23}-L^{32}$. (Note that similar definitions of the independent
Lee forms in the case of almost quaternionic manifolds are given
in~\cite{Ivq}.) One also notes the relation 
\bea
*(J^A\wedge J^B\wedge \dd J^C)=J^A\cdot L^{BC} + J^B\cdot L^{AC} .
\eea

Finally in later calculations we found it useful to determine the
relationships between the ten six-forms $J^A\wedge J^B\wedge J^C$.
A general six-form, which is Hodge-dual to a two-form,
corresponds to the $\Symp(2)$ representations in the
decomposition $\rep{28}\to
\rep{10}+3(\rep{5})+3(\rep{1})$. As the six-forms of interest are
constructed from $\Symp(2)$-singlets, they must correspond to
the three singlets in the decomposition, and hence there
must be seven relationships amongst the ten six-forms. They are
given by
\begin{equation}
\begin{gathered}
   J^1\wedge J^2\wedge J^2 
      = J^1\wedge J^3\wedge J^3
      = \tfrac{1}{3}J^1\wedge J^1\wedge J^1 , \\
   J^2\wedge J^3\wedge J^3
      = J^2\wedge J^1\wedge J^1
      = \tfrac{1}{3}J^2\wedge J^2\wedge J^2 , \\
   J^3\wedge J^1\wedge J^1
      = J^3\wedge J^2\wedge J^2
      = \tfrac{1}{3}J^3\wedge J^3\wedge J^3 , \\
   J^1\wedge J^2\wedge J^3 =0 .
\end{gathered}
\end{equation}

\smallsec{$\SU(2)\times\SU(2)$-structures in $d=8$}
The structure is defined by a pair of orthogonal
$\SU(2)$-structures which we can write as two triplets of almost
complex structures $(J^A,J'{}^A)$ satisfying  
\begin{equation}
\label{2SU2struc}
\begin{aligned}
   J^{A}\cdot J^{B} &=-\delta^{AB}\id +\epsilon^{ABC}J^{C} , \\
   J'{}^{A}\cdot J'{}^{B} &= -\delta^{AB}\id +\epsilon^{ABC}J'{}^{C} , \\
   J^{A}\cdot J'{}^{B} &= 0 .
\end{aligned}
\end{equation}
Again these define a metric. Lowering one index on the almost complex
structures gives six half-maximal rank two-forms. We also have a
natural eight-dimensional orientation given by $\vol\wedge\vol'$ where
$\vol=(J^A)^2/2$ and $\vol'=(J'{}^B)^2/2$ for any $A$ and
$B$.

Following the usual prescription decomposing the adjoint
representation of $\SO(8)$ into $\SU(2)\times\SU(2)$ representations
to give $(\su(2)\otimes\su(2))^\perp$ one finds 28 different real
modules: 
\begin{equation}
\begin{aligned}
   T \in \Lambda^1 \otimes (\su(2)\otimes\su(2))^\perp
      &= \bigoplus_{i=1}^{28} \mathcal{W}_i , \\
   ((\rep{2}+\rep{\bar{2}},\rep{1}) +(\rep{1},\rep{2}+\rep{\bar{2}})) 
      \times (6(\rep{1},\rep{1}) 
         + (\rep{2}+\rep{\bar{2}},\rep{2}+\rep{\bar{2}}))
      &= \\
         10(\rep{2}+\rep{\bar{2}},\rep{1}) 
         + 10(\rep{1},\rep{2}+\rep{\bar{2}}) 
         +& 4(\rep{3},\rep{2}+\rep{\bar{2}}) 
         + 4(\rep{2}+\rep{\bar{2}},\rep{3}) .
\end{aligned}
\end{equation}
Since the $\SU(2)$-structures are orthogonal, we necessarily have an
almost product structure $\Pi$. This is a tensor $\Pi_m{}^n$
satisfying $\Pi\cdot\Pi=\id$. It can be written in terms of the
complex structure as $\Pi=J^A\cdot J^A-J'{}^B\cdot J'{}^B$ for any $A$
and $B$. This can be written as the product of two commuting almost
complex structures $J^\pm=J^A\pm J'{}^B$. As discussed in
appendix~\ref{app:product}, generically the almost product structure
is not integrable.


\section{Geometries with $\epsilon^+$ Killing spinors in canonical
  dimension} 
\label{sec:canon}


We now consider generic supersymmetric type II geometries 
$(M_d,g_d,\Phi,H)$ when only one of the connections $\nabla^\pm$ 
has special holonomy. 
For definiteness we choose it to be $\nabla^+$. The different possible
holonomies are the usual groups given in figure~\ref{fig}. In this section
we will only consider geometries with $\nabla^+$ having special
holonomy in its minimal canonical dimension: the cases
are listed in table~\ref{green}. Our aim is to summarise the known cases
in a uniform way as well as to present new results on the two
remaining cases, $\Symp(2)$ and $\SU(2)\times\SU(2)$. At the end
of the section we will also discuss the generalisations needed 
for the heterotic/type I string theories. 

The basic technique to derive the results of this and subsequent sections
is to construct tensors from bi-linears in the Killing spinor $\e^+$, which
characterise the structure. Differential constraints on the structure
are obtained from the vanishing of the dilatino and gravitino variations.
The expression for the three-form $H$ as a generalised calibration, that
we are emphasising, can easily be obtained using the method of 
\cite{Gauntlett:2001ur}. We will not present any details of these
calculations in this section, for reasons of clarity. Note, however, that the
next section will contain some representative calculations.

\smallsec{$\SU(n)$-geometries in $d=2n$}
We start with the case where $\nabla^+$ has $\SU(n)$ holonomy in
$d=2n$ first considered in the case of heterotic/type I theories
in~\cite{strominger}. The necessary and sufficient conditions for
preservation of supersymmetry are that the manifold $M_{2n}$ has an
$SU(n)$ structure satisfying the differential 
conditions 
\begin{equation}
\label{SUncond}
\begin{aligned}
   \diff (\ex^{-2\Phi}\Omega) &= 0 , \\
   \diff (\ex^{-2\Phi} * J) &= 0 ,
\end{aligned}
\end{equation}
with the flux given in terms of the structure, in each case, by 
\cite{Gauntlett:2001ur}
\begin{equation}
\label{sunflux}
\begin{aligned}
   *H &= -\ex^{2\Phi} \diff (\ex^{-2\Phi}) 
      && \qquad \text{for $\SU(2)$} , \\ 
   *H &= -\ex^{2\Phi} \diff ( \ex^{-2\Phi} J ) 
      && \qquad \text{for $\SU(3)$} , \\
   *H &= -\ex^{2\Phi}\diff ( \ex^{-2\Phi}\tfrac{1}{2}J\wed J )
      && \qquad \text{for $\SU(4)$} .
\end{aligned}
\end{equation}
Note that here and throughout the paper the Hodge star is
defined with respect to the canonical orientation fixed by the
structure. For $\SU(n)$ this is $\vol=J^n/n!$. In terms of Killing
spinors, the geometries preserve two complex chiral $d=2n$ spinors
related by complex conjugation. For $n=2,4$ both spinors have the same
chirality, while for $n=3$ they have opposite chirality. Our
conventions for defining the spinors, $J$, $\Omega$ and $\vol$ are
given in Appendix~\ref{app:spinor}. 

These conditions on $J$ and $\Omega$ are equivalent to those
in~\cite{strominger} (after setting the gauge field to zero). In
particular, as we discuss below, they imply that $J$ is integrable. As
a result, the expression for $H$ can be rewritten in the form, as 
given in~\cite{strominger},
\begin{equation}
\label{strom}
   H = \ii (\bar \partial-\partial) J , 
\end{equation}
where $\diff=\bar \partial+\partial$. (Note that this corrects a sign
in the corresponding expression in~\cite{strominger}.\footnote{To see
  this one must take into account that our convention for the
  definition of $H$ has the opposite sign (and a factor of two) to that
  in~\cite{strominger}.}) However, it is the form~\p{sunflux} that
naturally generalises to other cases.

In particular we note that the expression for the three-form flux is
that of a generalised \Ka calibration. This is physically reasonable
since we expect geometries with flux should arise as solutions
describing fivebranes wrapping supersymmetric cycles, as discussed in
detail in \cite{Gauntlett:2001ur}. For instance, in
the $\SU(4)$ case, geometries with non-zero flux  with $\nabla^+$
having $\SU(4)$-holonomy correspond to a fivebrane wrapped on a \Ka
four-cycle in a Calabi-Yau four-fold. Such branes are calibrated by
$\frac{1}{2}J\wedge J$ which is precisely the generalised calibration
appearing in the expression for $H$. Similarly, the $\SU(3)$
geometries correspond to fivebranes wrapping \Ka two-cycles in $CY$
three-folds which are calibrated by $J$. The solutions found in
\cite{Maldacena:2000yy} are of this type (see
\cite{Papadopoulos:2000gj} for an explicit discussion). 
Finally the slightly degenerate $\SU(2)$ case corresponds to a
fivebrane wrapping a point in a $CY$ two-fold, i.e., the fivebrane is
transverse to the $CY_2$. Such configurations are calibrated by the
unit function.

The conditions on the $\SU(n)$ structure~\eqref{SUncond} can be rephrased
in terms of the classification of intrinsic torsion. The first
condition in~\p{SUncond} implies that $W_1=W_2=0$, and hence the almost complex
structure is in fact integrable  (as pointed out in~\cite{strominger}).
Thus for $\SU(3)$ and $\SU(4)$ the intrinsic torsion lies in
$\mathcal{W}_3\oplus\mathcal{W}_4\oplus\mathcal{W}_5$. For $\SU(2)$,
since $\mathcal{W}_1$ and $\mathcal{W}_3$ are always absent, we have
$T\in\mathcal{W}_4\oplus\mathcal{W}_5$. In all cases the second
condition in~\eqref{SUncond} is equivalent to the statement that
the Lee-form is exact and related to $\Phi$, namely $W_4=2\dd\Phi$. The
first condition also implies that Lee form for $\Omega$ is similarly
proportional to $\dd\Phi$ with $W_5=(-1)^n 2^{n-2} W_4$. For $\SU(3)$, 
this was first noticed in~\cite{dallagata}.

Note that this relation implies that under a conformal transformation,
the invariant combination~\eqref{confinv} is proportional to
$(n-2)W_4$. Thus only when $n=2$ is it possible to have 
geometries that are conformal to Calabi--Yau $n$-folds, as noticed by
\cite{strominger}. In this case $W_5=W_4=2\dd\Phi$ with $W_2=0$. The
general form of these geometries in ten dimensions is thus given by 
\begin{equation}
\label{NSinK3}
\begin{gathered}
   \diff s^2 =  
       \diff s^2 (\bbR^{1,5}) + \ex^{2\Phi}\diff \tilde{s}^2 , \\
   \tilde\nabla^2 \ex^{2\Phi} =  0~,
\end{gathered}
\end{equation}
with $H$ given as in~\p{sunflux} and $\dd\tilde{s}^2$ the metric on
$\CY_2$. This is is just the usual fivebrane solution transverse to
$\CY_2$. The possibility of conformally $\CY_2$ geometries was
considered in \cite{strominger} but here we claim the stronger result
that it is in fact necessary.   

It is worth emphasising that if $\Phi=\text{constant}$ then
the leading order equations of motion imply $H=0$ and in addition
$F=0$ for the heterotic/type I case (see for
example~\eqref{Phieom}). Thus, for instance, the solutions 
presented in \cite{dallagata} based on the Iwasawa manifold, although
supersymmetric, do not solve the leading order equations of
motion.  In general, solutions with $H\ne0$, and $\Phi$ non-constant
must have $W_4\ne0$ and $W_5\ne0$. Similar comments apply to other
cases considered below.  

\smallsec{\spin-geometries in $d=8$} 
Now consider the case when $\nabla^+$ has \spin holonomy.
The only condition on the \spin structure is that the Lee-form is
again exact \cite{Iv3}
\begin{equation}
\label{spincond}
   W_1 = 12 \dd\Phi
\end{equation}
with flux given by \cite{Gauntlett:2001ur}
\begin{equation}
\label{spinH}
   * H = -\ex^{2\Phi}\diff \left(\ex^{-2\Phi}\Psi\right) .
\end{equation}
These geometries preserve a single chiral spinor of $Spin(8)$.
As in the $\SU(n)$ case we can understand these geometries and
conditions in terms of wrapped branes. They arise as solutions for fivebranes
wrapping Cayley four-cycles in manifolds with \spin holonomy and the
expression for $H$ indeed corresponds to a generalised calibration for
such a cycle. 

It is interesting to note that if we perform a conformal transformation
$\tilde g\equiv \ex^{-6/7\Phi}g$, then the corresponding
$\Spin(7)$-structure defining $\tilde g$ has vanishing Lee-form, and 
hence has intrinsic torsion just in the class ${\cal W}_2$~\cite{Iv3}. 
One might entertain the idea of solutions that are conformal to a
\spin holonomy manifold, i.e. with  $\tilde g$ having \spin
holonomy. While such a geometry, with non-vanishing flux, certainly
admits Killing spinors, we cannot solve the Bianchi identity $\dd H=0$
with non-zero flux. To see this observe that the geometry has 
the form
\begin{equation}
\label{confspin7}
\begin{aligned}
   g &= \ex^{6/7\Phi}\tilde g , \\
   H_{mnp} &= 
      -\frac{1}{3}\tilde \Psi_{mnp}{}^q\tilde\nabla_q(\ex^{6/7\Phi}) .
\end{aligned}
\end{equation}
The expression for $\dd H$ contains both the ${\bf 35}$ and $\rep{1}$
representations of \spin. The singlet is proportional to
$\tilde\nabla^2(\ex^{6/7\phi})$ while the ${\bf 35}$ corresponds to the
trace-free part of  $\tilde\nabla_l\tilde\nabla_p(\ex^{6/7\phi})$.
We thus conclude
that $\dd H=0$ implies that $\Phi=\text{constant}$ which in turn
implies $H=0$.

\smallsec{$G_2$-geometries in $d=7$}
Next consider the case when $\nabla^+$ has $G_2$ holonomy. These geometries
preserve a single $d=7$ spinor. The
necessary conditions were derived
in~\cite{Gauntlett:2001ur,Friedrich:2001nh,Friedrich:2001yp} 
and sufficiency was proved in~\cite{Friedrich:2001nh,Friedrich:2001yp}.
This case was discussed in detail from the point of view of this
paper in~\cite{Gauntlett:2002sc}. The conditions placed on the
$G_2$ structure are given by 
\begin{equation}
\label{G2cond}
\begin{aligned}
   \phi \wedge \diff \phi &= 0 , \\
   \diff (\ex^{-2\Phi} * \phi) &= 0 ,
\end{aligned}
\end{equation}
which means that the intrinsic torsion lies in
$\mathcal{W}_3\oplus\mathcal{W}_4$ in the representations
$\rep{27}+\rep{7}$. Moreover it implies that the Lee form is again
exact with $W_4=-6\dd\Phi$. The flux is given by~\cite{Gauntlett:2001ur}
\begin{equation}
\label{G2flux}
   * H = \ex^{2\Phi}\diff (\ex^{-2\Phi}\phi) .
\end{equation}
It is worth noting that these geometries are special cases of
integrable $G_2$-structures in which one can introduce
a $G_2$ Dolbeault cohomology~\cite{fernug}.

These backgrounds arise
as solutions describing fivebranes wrapped on associative three-cycles in
manifolds of $G_2$ holonomy. This is reflected in the expression for
the flux which is the condition on a generalised calibration for such
a cycle. Solutions of this type were presented
in~\cite{Acharya:2000mu,Maldacena:2001pb,Gauntlett:2002sc} 
(see~\cite{Gauntlett:2002sc} for an explicit demonstration
of~\cite{Acharya:2000mu}).  

If we perform a conformal transformation $\tilde g\equiv \ex^{-\Phi}g$,
then the corresponding $G_2$-structure has vanishing Lee-form, and 
hence has intrinsic torsion just in the class
$\mathcal{W}_3$~\cite{Friedrich:2001yp}.  In particular one can
consider an ansatz for solutions that are conformal to a
$G_2$-holonomy manifold: 
\begin{equation}
\label{confg2}
\begin{aligned}
   g &= \ex^{\Phi}\tilde g , \\
   H_{mnp} &= -\tfrac{1}{2}\tilde *\tilde\phi_{mnp}{}^q
      \nabla_q(\ex^{\Phi}) .
\end{aligned}   
\end{equation}
However, as in the \spin case, \p{confspin7}, the Bianchi identity $\dd H=0$
implies that $\Phi$ is constant and hence $H=0$.

\smallsec{$\Symp(2)$-geometries in $d=8$}
Next consider the case when $\nabla^+$ has $\Symp(2)$ holonomy.
Such geometries are examples of manifolds known as hyper-K\"ahler 
with torsion. A discussion of these geometries can be found, 
for example, in~\cite{poon} and also~\cite{Ivq}. The dilaton
further constrains the geometry in the following way. The conditions
on the structure are given by those for the $SU(4)$ case for each
complex structure 
\begin{equation}
\label{sp2cond}
\begin{aligned}
   \diff(\ex^{-2\Phi}\Omega^{A}) &= 0 , \qquad &&\text{for $A=1,2,3$}, \\
   \diff(\ex^{-2\Phi} * J^{A}) &= 0 , \qquad&&\text{for $A=1,2,3$} ,
\end{aligned}
\end{equation}
with the flux being given by
\begin{equation}
\label{sp2H}
   * H = -\ex^{2\Phi}\diff \left(
         \ex^{-2\Phi}\tfrac{1}{2}J^{A}\wedge J^{A} \right) ,
      \qquad \text{for $A=1,2,3$} .
\end{equation}
These geometries preserve three chiral $d=8$ spinors with the same
chirality.
The conditions \p{sp2cond} imply that the parts of $\dd J^A$ transforming
in the two {\bf 16}'s are independent of $A$. In addition the 12 {\bf 4}'s
are determined by the dilaton. The ``diagonal'' Lee-forms
are all equal $L^{11}=L^{22}=L^{33}=2\dd\Phi$ and hence the off
diagonal Lee-forms $L^{AB}$, $A\ne B$ are anti-symmetric with
$L^{12}=-2J^3\cdot\dd\Phi$, $L^{31}=-2J^2\cdot\dd\Phi$ and 
$L^{23}=-2J^1\cdot\dd\Phi$.

It is worth noting that this case arises when fivebranes
wrap quaternionic planes in $\bbR^8$, that is cycles that are complex
with respect to all three complex structures. It was shown in
\cite{dadok} that these are linear. In \cite{george} solutions were
written down for these configurations and it is plausible that
they are the most general, once the Bianchi identity is imposed.

\smallsec{$SU(2)\times SU(2)$-geometries in $d=8$}
Finally consider the case when $\nabla^+$ has $SU(2)\times SU(2)$
holonomy. The conditions on the structure are
\begin{equation}
\label{susucond}
\begin{aligned}
   \dd(\ex^{-2\Phi}J^A\wedge \vol') &=0 , \\
   \dd(\ex^{-2\Phi}J^A{}'\wedge \vol)&=0 , \\
   \dd(\ex^{-2\Phi}J^A\wedge J^B{}')&=0 ,
\end{aligned}
\end{equation}
where, e.g., $\vol=(J^A\wedge J^A)/2$ for each $A$,
while the flux is given by
\begin{equation}
\label{susuH}
   *H = -\ex^{2\Phi}\diff \left(\ex^{-2\Phi}\vol+\ex^{-2\Phi}\vol'\right) .
\end{equation}

These geometries preserve four chiral $d=8$ spinors, all with the same 
chirality. As discussed in Appendix~\ref{app:product}, the almost
product structure defined by $\Pi=(J^A+J'{}^B)\cdot(J^A-J'{}^B)$ is
not integrable. This is because the mixed components $H_{ija}$ and $H_{abi}$,
using the notation of Appendix~\ref{app:product}, are generically
non-zero. A notable subclass of solutions, with integrable products,
is given by those corresponding to two orthogonal fivebranes
intersecting in a string, one fivebrane wrapping $CY_2$ and the other
$CY_2'$ in $CY_2\times CY_2'$. Such solutions are discussed for
instance in~\cite{Gauntlett:1997pk}.   

\medskip

Let us now consider the modifications required for
heterotic/type I string theory. In addition to 
$g_d,H,\Phi$, the bosonic field content also includes a gauge field $A$, 
with field strength $F$, in the adjoint of $E_8\times E_8$ or $SO(32)/\bbZ_2$.
In order to preserve supersymmetry we require the expressions 
in \p{susy} for $\e^+$ only, and thus the cases described in table~\ref{green}
and the above discussion 
are equally applicable to the heterotic/type I theories.
In addition, preservation of supersymmetry requires 
the vanishing of the gaugino variation~\eqref{gaugino}
\bea
\Gamma^{MN} F_{MN}\e^+=0 .
\eea
{}For each case in table~\ref{green}, 
since $\e^+$ is a singlet of the special holonomy group $G$ of $\nabla^+$
this is satisfied, breaking no further supersymmetry, if the
two-form $F$, considered as the adjoint of $SO(d)$, 
lies within the adjoint of $G$.

{}For the \spin case we therefore need to consider $F$ to be a \spin
instanton satisfying
\begin{equation}
   F_{mn}=-\tfrac{1}{2} \Psi_{mn}{}^{pq}F_{pq}~,
\end{equation}
while for $G_2$ we need
\begin{equation}
   F_{mn}=-\tfrac{1}{2} *\phi_{mn}{}^{pq}F_{pq}.
\end{equation}
{}For the $\SU(n)$ cases, we require
\begin{equation}
\label{spgfcon}
   F_{mn}=-\tfrac{1}{2} \left(\tfrac{1}{2}J\wed J\right)_{mn}{}^{pq}F_{pq}   
\end{equation}
which, in complex coordinates, is equivalent to
\begin{equation}
   J^{\alpha\bar \beta}F_{\alpha\bar \beta}
      = F_{\alpha\beta} = F_{\bar \alpha \bar \beta} = 0 .
\end{equation}
That is we need a holomorphic gauge field on a holomorphic vector bundle
satisfying the Donaldson-Uhlenbeck-Yau equation, as noticed in 
\cite{strominger}. For the $\Symp(2)$-case we require that the gauge 
field satisfies \p{spgfcon} for all three complex structures, or equivalently,
\begin{equation}
   F_{mn} = J^A_m{}^p J^A_n{}^q F_{pq},\qquad \text{no sum on $A$},
\end{equation}
which are the same as the BPS equations of \cite{ward}. For
$\SU(2)^2$, with self-dual complex structures,
the gauge fields must describe an anti-self-dual 
instanton
for each of the $\SU(2)$ structures. This can be written as
\begin{equation}
   F_{mn} = - \tfrac{1}{2}\vol_{mn}{}^{pq} F_{pq}
      = - \tfrac{1}{2}\vol'_{mn}{}^{pq} F_{pq} .
\end{equation}
Note that in all case the instanton condition can be written as 
\begin{equation}
\label{eq:instanton}
   * F = \Xi \wedge F
\end{equation}
where $\Xi$ is the invariant form entering the generalised calibration
expression for the flux $*H=\ex^{2\Phi}\dd(\ex^{-2\Phi}\Xi)$. 

As shown in \cite{Gauntlett:2002sc} the equations of motion of 
type I supergravity are automatically satisfied if one imposes 
the modified Bianchi identity for $H$
\bea
\label{eq:BI1}
\dd H = 2\alpha' \Tr F\wedge F .
\eea
In type I/heterotic string theory the Bianchi identity is modified by
higher order corrections 
\bea
\label{eq:BI2}
\dd H = 2\alpha'(\Tr F\wedge F - \tr R\wedge R)
\eea
which allows solutions with $\dd H=0$ as for the type II theories.

We noted above for the \spin case
that the ansatz~\p{confspin7} preserves Killing spinors
but does not solve the Bianchi identity $\dd H=0$, and hence the equations
of motion, for non-vanishing $H,\Phi$. It is interesting to ask whether
there are heterotic solutions solving $\dd H=2\alpha'\Tr F\wedge F$.
Indeed, when $\tilde g$ is flat such solutions have already been
found~\cite{harvstrominger}. Similarly heterotic solutions for $d=7$
that are conformal to flat space were found in
\cite{gunnicolai}. It would be interesting to construct heterotic
solutions when $\tilde g$ is conformal to a non-flat \spin or $G_2$-holonomy
manifold.


\section{General geometries with $\epsilon^+$ Killing spinors}
\label{sec:gen}


In the previous section, we gave the necessary and sufficient
conditions for preservation of supersymmetry for a geometry of the
form $\bbR^{1,9-d}\times M_d$ when $\nabla^+$ has special holonomy in
the corresponding canonical number of dimensions, $\Spin(7)$ in $d=8$,
$G_2$ in $d=7$ and so on. The analysis for
$\nabla^-$ is simply obtained by taking $H\to -H$. 
More generally one can ask for the generic static
supersymmetric background of the form $\bbR\times M_9$ preserving some
number of supersymmetries.  In this section, we give a complete
analysis of this question when the spinors are all of the same type 
and show that in addition to recovering the results of the previous
section we find more general classes of geometries. As before, for
definiteness we take the Killing spinors to be all of the type
$\epsilon^+$ satisfying $\nabla^+\epsilon^+=0$. In the next section we
turn to the case where some Killing spinors satisfy
$\nabla^+\epsilon^+=0$ and some $\nabla^-\epsilon^-=0$.

Suppose we have $N$ independent spinors $\epsilon^+_{(i)}$ in $d=9$
all satisfying $\nabla^+\epsilon^+_{(i)}=0$. In general, these define
a $G$-structure, where $G\subset\Spin(9)$ is the stabiliser group of
rotations which leave all the spinors invariant. One finds the
seven special holonomy groups given in
figure~\ref{fig} as possibilities. Furthermore these embed in
$\Spin(9)$ in the conventional way following the pattern of the
dimensional reduction. That is to say $G\subset\SO(n)\subset\SO(9)$
where $n$ is the canonical dimension for the $G$-structure as given in
figure~\ref{fig}. 

As usual the structures can also be defined in terms of a set of forms
which can be constructed out of the spinors. In general, these are of
the type $(K^1,\dots,K^{9-n},\Xi^A)$ with $i_{K^i}\Xi^A=0$. Here
$\Xi^A$ are the set of forms used to define the structure in its
canonical dimension $n$ as described in
section~\ref{sec:Gstructure}. The $K^i$ are a
set of $9-n$ independent
one-forms required to define the additional orthogonal dimensions to
give a structure in $d=9$. Thus for instance a $G_2$-structure in $d=9$
is defined by the set $(K^1,K^2,\phi)$ with $i_{K^i}\phi=0$. In a local
orthonormal frame $e^m$, we can take the form $\phi$ to have the
standard form~\eqref{Psidef} in terms of $e^1,\dots,e^7$ while
$K^1=e^8$ and $K^2=e^9$. Thus, at any given point in $M_9$, the forms
$K^1$ and $K^2$ define a reduction of $\bbR^9$ into
$\bbR^7\oplus\bbR^2$ and hence define a $\SO(7)\subset\SO(9)$
structure. The three-form $\phi$ then describes a $G_2\subset\SO(7)$
structure on the $\bbR^7$ subspace in the usual way. Note that the
structure always defines a metric. Using this metric we can also view
the $K^i$ as vectors, which we will also denote as $K^{i\#}$. 
In addition, as we will see, the inner product
$K^i\cdot K^j$ is constant for all $i$ and $j$ and so we normalise
$K^i$ to be orthonormal.  

If the flux $H$ is zero, we have $\nabla K^i=0$ and $M_9$ is then, after
going to the covering space, just
a product $M_9=\bbR^{9-n}\times M_n$ where $M_n$ is a
$G$-holonomy manifold in the canonical dimension. From this point of
view, $G$-holonomy extends trivially to nine dimensions. With flux
however, this is no longer the case. We will show that there are new
possibilities which are not simply direct products of the geometries
given in the 
previous section with flat space. We discuss the most general
case of $G=\Spin(7)$, corresponding to one Killing spinor, 
in detail and then summarise the analogous results for the other
structure groups, corresponding to the existence of more than one
Killing spinor.

\subsection{Single Killing spinor: $\Spin(7)$-structure in $d=9$}
\label{sec:d=9}

First assume we have a single Killing spinor $\epsilon^+$ on $M_9$, and
since $\nabla^+\epsilon^+=0$, we can take $\bar\e^+\e^+=1$. 
It is easy to show that the stability group is
$\Spin(7)\subset\Spin(9)$. Equivalently we have the set of
$\Spin(7)$-invariant forms $(K,\Psi)$ with $i_K\Psi=0$ and $K^2=1$. In a
particular basis $e^m$, we can take $K=e^9$ and $\Psi$ given by the
standard form~\eqref{Psidef} in terms of $e^1,\dots,e^8$. In terms of
the spinor $\epsilon^+$, we have 
\begin{equation}
   K_m = \bar{\epsilon}^+ \gamma_{m} \epsilon^+ , \qquad
   \Psi_{mnpq} = -\bar{\epsilon}^+ \gamma_{mnpq} \epsilon^+ ,
\end{equation}
where $\gamma_m$ are nine-dimensional gamma matrices with
$\gamma_{1\cdots9}=\id$. From the Killing spinor
conditions~\eqref{susy}, as in the previous section, one derives a set
of necessary and sufficient conditions on $(K,\Psi)$. The condition
$\nabla^+\epsilon^+=0$ simply translates into
$\nabla^+\Psi=\nabla^+K=0$. From the latter constraint we immediately
see, since $H$ is totally antisymmetric, that $K$ is a Killing vector,
and in addition that the norm of $K$ is constant, as claimed above. In
addition one finds 
\begin{equation}
\label{dK}
   \dd K = G ,
\end{equation}
where we have made the generic $\SO(8)$ decomposition  
\begin{equation}\label{9dthree}
   H \equiv H_0 - K \wedge G ,
\end{equation}
with $i_KH_0=i_KG=0$. We can now introduce local
coordinates such that the metric has the canonical form of a fibration 
\begin{equation}
\label{9dmetric}
   \dd s^2 = \dd s^2(M_0) + (\dd y+B)^2 ,
\end{equation}
with $K=\dd y+B$, while $\dd B=G$ is a two-form on $M_0$ and the
metric $\dd s^2(M_0)$ is independent of $y$ and admits a
$\Spin(7)$-structure defined by $\Psi$, which may, however, at this point,
depend on $y$. 

Now we turn to the dilatino equation. Following the discussion 
in~\cite{Gauntlett:2001ur}, given the symmetry properties of the 
nine-dimensional gamma matrices, one has 
\begin{equation}
\label{eq:commutator}
   \partial_m\Phi {\bar \epsilon^+}[A,\gamma^m]_\mp\epsilon^+
      + \frac{1}{12}H_{mnp}
          {\bar\epsilon^+}[A,\gamma^{mnp}]_\pm\epsilon^+
      = 0 
\end{equation}
where $A$ is an operator built out of gamma matrices and
$[\,\cdot\,,\,\cdot\,]_\pm$ refer to the anti-commutator and
commutator respectively. By taking $A=\gamma^{m_1}$ with the lower
sign and $A=\gamma^{m_1\dots m_6}$ with the upper sign
in~\eqref{eq:commutator}, one finds two constraints on
$(K,\Psi)$. First one has the Lee-form condition
\begin{equation}
\label{spinsevenlee}
   \Psi\lrcorner\, \dd\Psi = 12\dd\Phi , 
\end{equation}
and then the familiar calibration form for the flux 
\begin{equation}
\label{9dflux}
   *H = \ex^{2\Phi}\dd ( \ex^{-2\Phi}\Psi\wedge K ) .  
\end{equation}
Note that we have fixed our orientation by $\vol_{m_1\dots
  m_9}=\bar{\epsilon}^+\gamma^{m_1\dots m_9}\epsilon^+$. 

If we decompose ~\eqref{9dflux} into $\SO(8)$ representations, consistency
with~\eqref{dK} requires
\begin{equation}
   G_{mn} = -\tfrac{1}{2}\Psi_{mn}{}^{pq}G_{pq} . 
\end{equation}
In other words, $G$ satisfies the $\Spin(7)$ instanton equation on
$M_0$. As a result, $K$ is not only a Killing vector but actually
preserves the $\Spin(7)$ structure. That is, the Lie-derivative of 
the spinor $\e^+$ vanishes and hence the Lie derivative
of $\Psi$ also vanishes,   
\begin{equation}
   \mathcal{L}_K\Psi = 0 ,
\end{equation}
which implies
similarly that $\mathcal{L}_KH=\mathcal{L}_K\Phi=0$. The Lee-form 
condition in~\eqref{spinsevenlee} can then be written 
\begin{equation}\label{oldlee}
   \Psi\lrcorner\, \dd_0\Psi = 12\dd_0\Phi
\end{equation}
where $\dd_0$ is the exterior derivative on the eight-dimensional
space $M_0$. Similarly the condition~\eqref{9dflux} reduces to 
\begin{equation}\label{oldfluxcond}
   *_0 H_0 = -\ex^{2\Phi}\dd_0 ( \ex^{-2\Phi}\Psi ) , 
\end{equation}
where $*_0$ is the Hodge-star on $M_0$. In other words, the $d=8$
\spin structure $\Psi$ on $M_0$ is independent of $y$ and satisfies
exactly the same conditions~\eqref{spincond} and~\eqref{spinH} as in
the last section. In particular, the only constraint
on the intrinsic torsion in $d=8$ is that the Lee form is
given as in~\eqref{spinsevenlee}. By substituting back into the supersymmetry
conditions \p{susy} it easy to see that these conditions are sufficient
for supersymmetry. We should point out that it is straightforward to
also define and characterise the intrinsic torsion of the \spin
structure directly in $d=9$ but as it provides no extra information on
how to characterise the geometries we shall not present any details here. 

To summarise, the general $d=9$ geometry is simply a flat direction 
fibred over a $d=8$ \spin geometry, with the fibration determined
by an Abelian \spin instanton in $d=8$. The metric is given by
\p{9dmetric}, the three-form by~\p{9dthree},~\p{oldfluxcond} and the
dilaton by~\p{oldlee}. In order to obtain a supersymmetric
solution to the equations of motion we also need to impose the
Bianchi identity for $H$. Explicitly we get
\begin{equation}
   \dd_0H_0 - G\wedge G = \begin{cases} 
      0 & \text{for type II} \\ 
      2\alpha' (\Tr F\wedge F -\tr R\wedge R)& \text{for heterotic/type I} 
      \end{cases} ,
\end{equation}
where $F$ is a \spin instanton.

A number of further comments are in order. First, when the flux is
zero, we commented above that, after going to the covering space,
the geometry is necessarily a direct product of a $d=8$ \spin
holonomy manifold  with a flat direction. By contrast when the 
flux is non-zero, it is only in the special
case when $\dd K=G=0$, when the fibration is trivial, that the
geometries are simply the product of the $d=8$ \spin geometries
considered in the last section with a flat direction. 

Secondly, since $K$ generates a symmetry of the full solution, 
including the spinors, we can dimensionally reduce a type II 
solution to get a supersymmetric heterotic solution in $d=8$ with an Abelian
instanton $F$ proportional to $G$. Similarly, given a heterotic solution
$(g_0,H_0,\Phi,F)$ in $d=8$ with an Abelian \spin instanton $F$, we
can oxidise it to obtain a type II solution in $d=9$ with $G$ proportional
to $F$, a metric given by~\eqref{9dmetric} and $H=H_0-G\wedge K$.  

Thirdly, the solutions are invariant under a $T$-duality in the
$y$-direction.

Finally, note that the $d=9$ expression for the flux \p{9dflux}
is again that of a
generalised calibration. It corresponds to a NS fivebrane wrapping a 
supersymmetric five-cycle $\Sigma_4\times S^1$ in the product of
a \spin manifold $\bar{M}$ with a circle, $\bar{M}\times S^1$, with
$\Sigma_4\subset\bar{M}$ being a Cayley four-cycle. (Note one could
equally well replace the circle with a line.) The simplest way of
wrapping the fivebrane leads to a $d=9$ geometry consisting of the
product of a $d=8$ \spin geometry considered in the last section with
a $S^1$. The $S^1$ is a flat direction on the world-volume of the
fivebrane. The analysis of this section shows that more complicated
geometries can arise leading to the world-volume direction being
fibred over the $d=8$ manifold. As wrapped branes have holographic
duals, it will be interesting to determine the holographic
interpretation of this.

\subsection{Multiple Killing spinors}

The case of multiple $\e^+$ Killing spinors is completely analogous to the
$\Spin(7)$ case discussed above. As mentioned, the set of spinors
$\epsilon^+_{(i)}$ in general define a $G$-structure in $d=9$ with
$G$ being one of the standard special holonomy groups $\SU(4)$,
$\Symp(2)$, $\SU(2)\times\SU(2)$, $G_2$, $\SU(3)$ or $\SU(2)$.
 One way to view how these groups appear is to see
that the stability group of each $\epsilon^+_{(i)}$ defines a
different embedding of $\Spin(7)$ in $\Spin(9)$. The structure group
$G$ is then the common subgroup of this set of embedded $\Spin(7)$
groups. From this perspective, each $G$-structure is equivalent to a
set of distinct $\Spin(7)$-structures.

The structure group $G$ is then the common subgroup of these
embeddings. Recall the structure can be defined in terms of
$(K^i,\Xi^A)$ where $\Xi^A$ are forms used to define the structure in its
canonical dimension $n$ and $K^i$ are $9-n$ one-forms. The condition
$\nabla^+K^i=0$ implies each $K^i$ is Killing and we can take them
to be orthonormal. In addition, as in the
$\Spin(7)$ case one can always derive as set of necessary and
sufficient conditions on $(K^i,\Xi^A)$ using the dilatino
constraint. One always finds the familiar calibration
condition for $*H$. Explicitly, for the cases where $n=8$ one has  
\begin{equation}
\label{calib1}
   * H = \begin{cases} 
      \ex^{2\Phi} \dd \left( \ex^{-2\Phi}
          \tfrac{1}{2}J\wedge J \wedge K \right) 
             & \text{for $\SU(4)$} , \\ 
      \ex^{2\Phi} \dd \left( \ex^{-2\Phi}
          \tfrac{1}{2}J^A\wedge J^A \wedge K \right) 
             & \text{for $\Symp(2)$ with $A=1,2,3$} , \\ 
      \ex^{2\Phi} \dd \left( \ex^{-2\Phi}\vol\wedge K
          + \ex^{-2\Phi}\vol'\wedge K \right) 
             & \text{for $\SU(2)\times\SU(2)$} , \\ 
      \end{cases}
\end{equation}
where $K$ is the single one-form, while for the $n<8$ cases we have
\begin{equation}
\label{calib2}
   * H = \begin{cases} 
      \ex^{2\Phi} \dd \left( \ex^{-2\Phi}
          \phi \wedge K^1 \wedge K^2 \right) 
             & \text{for $G_2$} , \\ 
      \ex^{2\Phi} \dd \left( \ex^{-2\Phi}
          J\wedge K^1 \wedge K^2 \wedge K^3 \right) 
             & \text{for $\SU(3)$} , \\ 
      \ex^{2\Phi} \dd \left( \ex^{-2\Phi}
          K^1 \wedge \dots \wedge K^5 \right) ,
             & \text{for $\SU(2)$} . \\ 
      \end{cases}
\end{equation}

The necessary and sufficient conditions also imply that the Killing vectors
$K^i$ all commute and furthermore each preserves the underlying
$G$-structure $\Xi^A$. This implies that the metric can be put in the
canonical fibration form    
\begin{equation}
   \dd s^2 = \dd s^2(M_0) + \sum_{i=1}^{9-n}(\dd y^i+B^i)^2 ,
\end{equation}
where $M_0$ is a $n$-dimensional manifold and $K^i=\dd
y^i+B^i$. Furthermore, $M_0$ has  a $G$-structure defined by $\Xi^A$
independent of $y^i$. The flux $H$ has the related decomposition   
\begin{equation}
\label{Hexp}
   H \equiv H_0 - \sum_{i=1}^{9-n} K^i \wedge G^i ,
\end{equation}
where $G^i=\dd B^i$ are two-forms on $M_0$. In addition one finds a set
of constraints on the $G$-structure $\Xi^A$ on $M_0$. As in the $\Spin(7)$
case these turn out to be precisely the canonical dimension conditions
given in the last section. 

The additional freedom in nine-dimensional geometries are 
given by the two-forms
$G^i$ defining the fibration. Again as in the $\Spin(7)$ case
consistency between the calibration conditions~\eqref{calib1}
and~\eqref{calib2} and the expansion~\eqref{Hexp} implies that each
$G^i$ satisfies the appropriate Abelian $G$-instanton equation on
$M_0$. 

In summary, general supersymmetric geometries in $d=9$ are closely
related to the supersymmetric geometries in the canonical dimensions
discussed in the last section. They all have a fibred structure where
the base space $M_0$ has a $G$-structure in canonical dimension
satisfying one of the sets of conditions given in
section~\ref{sec:canon}. The flux is given by a generalised
calibration condition~\eqref{calib1} or~\eqref{calib2}, corresponding
to a fivebrane wrapping a five-cycle. The twisting of the fibration is
described by two-forms $G^i$ which are all Abelian $G$-instantons on
$M_0$. If one makes a dimensional reduction on the $K^i$, the
solutions correspond to heterotic solutions in canonical dimension
$d=n$ with $9-n$ Abelian instantons. 
In order to obtain a solution to the equations of motion
the flux $H_0$ on $M_0$ must also satisfy a modified Bianchi identity
\begin{equation}
   \dd_0 H_0 - \sum_{i=1}^{9-n} G^i \wedge G^i = \begin{cases} 
         0 & \text{for type II}, \\
         2\alpha'(\Tr F\wedge F -\tr R\wedge R)& \text{for heterotic/type I}. 
      \end{cases}
\end{equation}

These results provide a comprehensive
classification of all the possible
supersymmetric heterotic/type I or NS--NS type II 
bosonic geometries of the form
$\bbR^{1,9-d}\times M_d$ preserving Killing spinors satisfying
\p{susy} for $\e^+$. Any solution with $d<9$ can be obtained simply
by setting $9-d$ of the $B^i$ twists to zero, so that the fibration
becomes, at least partially, a product $M_9=\bbR^{9-d}\times M_d$. 

Note that there is one possible caveat to this analysis which is the
existence of geometries with exactly five, six or seven Killing
spinors. This necessarily defines an $\SU(2)$-structure and would require the
existence of a compatible connection $\nabla^+$ without the particular
fibration structure described in the text. Similar comments apply to
the existence of solutions with nine or more supersymmetries (so 
defining an identity structure) which are not simply flat space.  

It is interesting to note that 
particular examples of these general types of solutions have already
appeared in the literature. Examples of $SU(2)$-structure in $d=6$ and
$SU(2)^2$ in $d=9$ were considered in~\cite{Lu:2002xa} using
conformally Eguchi--Hanson metrics. Similar solutions related to
D3-branes were considered in \cite{Lu:2002rk}. Further examples will
be presented in the next section.

We should also note that $d=6$ geometries of the type
discussed here with two flat directions are similar to those studied in 
\cite{Goldstein:2002pg}. However, the motivation of that work
was rather different. Namely, the idea was to exploit
the fibration structure in order to construct examples of manifolds 
with $SU(3)$ structures in six-dimensions of the type described in the 
last section.


\section{Explicit examples I}
\label{sec:ex1}


We now present explicit solutions of the type described in the last
section. For illustration we shall consider here just a single
flat direction fibred over a base-manifold $M_0$. Additional examples
with two flat directions fibred over a four-dimensional base will be
considered in section~\ref{sec:ex2}. To begin with we consider
$M_0$ to be four-dimensional, and the three complex structures are
taken to be self-dual.  As noted in section~\ref{sec:canon}, $M_0$ is
necessarily conformally hyper-K\"ahler. The five-dimensional 
geometry thus takes the form
\begin{equation}
\label{explictd=5ex}
\begin{aligned}
   \dd s^2 &= \ex^{2\Phi}(\dd\tilde s^2)+(\dd y+B)^2 , \\
   H_{mnp} &= -\tilde \epsilon_{mnp}{}^l\tilde{\nabla}_l
      \ex^{2\Phi}-3B_{[m}G_{np]} , \\ 
   H_{ymn} &= -G_{mn} ,
\end{aligned}
\end{equation}
where $G=\dd B$ is an Abelian anti-self-dual instanton and
$\tilde{\epsilon}$ is the volume form on $\dd\tilde{s}^2$.
Generically, these solutions preserve 1/2 of the $\e^+$ supersymmetries,
and none of the $\e^-$ supersymmetries for the type II theories, 
corresponding to eight supercharges for both the heterotic and the
type II theories. For solutions, we must impose the Bianchi identity
for $H$. This gives 
\begin{equation}
\label{HBI}
   - \,\tilde{\dd}\,\tilde{*}\,\tilde{\dd}\,\ex^{2\Phi} 
      = \begin{cases} 
            G \wedge G & \text{for type II}, \\
            G \wedge G + 2\alpha'(\Tr F\wedge F -\tr R\wedge R)& 
               \text{for heterotic/type I}. 
         \end{cases} 
\end{equation}
Recall that supersymmetry implies that $F$ is also an anti-self-dual
instanton on the base. In the special case that
$\tr R\wedge R=0$, satisfying the Bianchi identity then implies that
the leading equations of motion are automatically
satisfied. Otherwise, one must separately check that one has a
solution of the equations of motion, including at this order $\alpha'$
corrections.  

Particular solutions can be found whenever we have an explicit
anti-self-dual Abelian instanton $G$ on a hyper-K\"{a}hler manifold. The
simplest cases are when the hyper-K\"ahler metric is flat. Let us present
some examples just for the type II case, for simplicity, where the 
Bianchi identity becomes
\begin{equation}
   \tilde\nabla^2\ex^{2\Phi}=-\tfrac{1}{2}\tilde G^2 . 
\end{equation}
Then a simple anti-self-dual instantons is given for instance by 
\bea
   B=\gamma(x^1 \dd x^2-x^3 \dd x^4),
\eea
corresponding to a constant field-strength. A radial solution for the
dilaton is given by 
\bea
   \ex^{2\Phi} = 1+\frac{m}{r^2} - \frac{1}{4}\gamma^2 r^2 .
\eea

A different radial solution can be obtained by writing the flat
metric in terms of left-invariant one-forms on the three-sphere:
\bea
   \dd s^2 = \dd r^2 
      + \tfrac{1}{4}r^2\left( 
         (\sigma^1_R)^2 + (\sigma^2_R)^2+ (\sigma^3_R)^2 \right)
\eea
with positive orientation given by $\dd r\wedge\sigma^1_R 
\wedge\sigma^2_R\wedge\sigma^3_R$ (our conventions are as in \cite{the5}).
A singular anti-self-dual instanton is then given by
\bea
B=\frac{\gamma}{4r^2}\sigma^3_R .
\eea
A radial solution for the dilaton is 
\bea
\ex^{2\Phi}=1+\frac{m}{r^2} -\frac{\gamma^2}{12 r^6} .
\eea

When the hyper-K\"ahler metric is Eguchi-Hanson space or Taub-NUT
space any of the anti-self-dual harmonic two-forms on these spaces can
be used as the Abelian instanton and if they are normalisable they
lead to non-singular solutions. These cases have already appeared in
the literature~\cite{Lu:2002xa}. 

Let us now consider whether we can obtain compact heterotic
solutions of the form~ \p{explictd=5ex}. (Recall that there are no
compact solutions with flux for the type II cases \cite{dWSHD}.) 
The base space
$\tilde{M}$ must admit a hyper-K\"ahler metric so is either $T^4$ or
$K3$. In addition, we will compactify the fibre direction on a circle $S^1$
of radius $R$. By construction such a background preserves eight 
supersymmetries. For a solution we must also satisfy the Bianchi
identity. The left-hand side of~\eqref{HBI} is exact, thus the sum of
the sources on the right-hand side must be trivial in
cohomology. Since the manifold is compact, each of the sources is
also quantised, being some multiple of the first Pontrjagin $p_1\in
H^2(M_5,\bbZ)$ class (instanton charge) of the corresponding
bundle. If $E$ is the bundle describing the $S^1$ fibration and $V$
the bundle of the heterotic/type~I gauge fields we have 
\begin{equation}
\label{quant}
   R^2 p_1(E) + 2\alpha' p_1(V) - 2\alpha' p_1(TM_5) = 0 
\end{equation}
in cohomology. Note that given the definition of $G$ the field strength
entering $p_1(E)$ is $G/R$ hence the factor of $R^2$ in the first
term. Since both $G$ and $F$ are anti-self-dual instantons on the base
$p_1(V)$ cannot cancel against $p_1(E)$ and we can only
satisfy~\eqref{quant} by including non-trivial $p_1(TM_5)$. The equation
for the dilaton on $\tilde{M}$ then becomes,  
\begin{equation}
   \tilde\nabla^2\ex^{2\Phi} = -\tfrac{1}{2}\tilde G^2  
     - \alpha'\left( \Tr \tilde F^2 - \tr \tilde R^2\right) .
\end{equation}
One would then have to check whether such a solution for $\Phi$ in
fact leads to a background satisfying the full (higher-order)
equations for motion. One important point to note is that
satisfying~\eqref{quant} with non-vanishing $p_1(E)$ requires
$R^2\sim\alpha'$. In other words the size of the the $S^1$ fibre must be
of order the string scale. As such the supergravity description of
these compactifications is breaking down. (Note, in addition, that
$R^2$ is constrained to be a rational multiple of $\alpha'$, so cannot
be a modulus.) It would be interesting to find a corresponding
conformal field theory description, for instance by taking the
orbifold limit of the base $K3$ manifold. Note that it is trivial to
extend these solutions to six-dimensional compactifications with
$\mathcal{N}=2$ supersymmetry simply by including a second fibred
direction.

Now let us consider solutions where the base geometry $M_0$ is in more 
than four dimensions. Specifically we consider solutions where $M_0$
is conformal to a special holonomy manifold. We noted in
section~\ref{sec:canon} that this rules out the $\SU(n)$-cases for
$n\ne 2$. Let us thus consider $M_0$ to be conformal to a
$G_2$-holonomy manifold. An eight dimensional geometry preserving two
$\e^+$ supersymmetries, one of each $d=8$ chirality, is given by
\begin{equation}
\begin{aligned}
   \dd s^2 &= \ex^{\Phi}(\dd\tilde s^2)+(\dd y+B)^2 , \\
   H_{mnp} &= -\frac{1}{2}\tilde * \tilde \phi_{mnp}{}^q
      \tilde{\nabla}_q \ex^{\Phi}-3B_{[m}G_{np]} , \\
   H_{ymn} &= -G_{mn}
\end{aligned}
\end{equation}
where $G=\dd B$ is an Abelian $G_2$-instanton on the $G_2$-holonomy
manifold $\tilde M$. A type II solution is then obtained by solving
the Bianchi identity which reads 
\bea
   \tilde * \tilde \phi_{[mnp}{}^q
      \tilde\nabla_{l]}\tilde \nabla_q\ex^{\Phi} = 3G_{[mn} G_{pl]} .
\eea
Given that $G$ is a $G_2$-instanton, this is equivalent to
\bea\label{dilat}
   \tilde \nabla_{m} \tilde \nabla_n\ex^\Phi = 
      -2\tilde G_{m}{}^kG_{nk}+ \tfrac{1}{4}\tilde G^2 \tilde g_{mn} . 
\eea

To get explicit solutions we need explicit $G_2$-holonomy metrics
$\dd\tilde{s}^2$ and explicit Abelian instantons $G$. One approach is
to note the if the $G_2$-holonomy metric admits a Killing vector $v$,
then the two-form $\dd v$ is a $G_2$-instanton if and only if $v$
preserves the $G_2$-structure: ${\cal L}_v\phi=\dd\,i_v\phi=0$. Since
all of the known explicit $G_2$-manifolds have many isometries, this
result allows one in principle to find new solutions and would be
interesting to investigate further.  

If the $G_2$-holonomy manifold is flat, solutions with constant flux
can be obtained as follows. We take 
\begin{equation}
   B = \frac{1}{2}C_{mn} x^m \dd x^n , 
\end{equation}
giving constant field strength $G=C$. This is an $G_2$ Abelian
instanton provided
$C_{mn}=-\frac{1}{2}\tilde{*}\tilde{\phi}_{mn}{}^{pq}C_{pq}$. In other
words, using a suitable projection, we have in general
\begin{equation}
    C_{mn} = \frac{2}{3}\left(
       \delta_{mn}^{pq}-\frac{1}{4}\tilde *\tilde\phi_{mn}{}^{pq}
       \right) D_{pq} ,
\end{equation}
for an arbitrary constant two-form $D_{mn}$. 
We then find that 
\begin{equation}
   \ex^\Phi = -\frac{1}{2}\left(
         2\tilde G_{m}{}^kG_{nk}-\frac{1}{4}\tilde G^2 \tilde g_{mn}
         \right)x^m x^n
      + \text{constant} 
\end{equation}
solves~\p{dilat}.


\section{Geometries with both $\e^+$ and $\e^-$ Killing spinors}
\label{sec:+-}


Let us now turn our attention to the type II cases summarised in
table~\ref{blue}. These geometries preserve both $\e^+$ and $\e^-$
Killing spinors and thus define two different structures, $G^\pm$, one
for each set of Killing spinors, of the type described in
section~\ref{sec:canon}. Taking both sets together defines a
$G$-structure where $G$ is the maximal common subgroup of $G^+$ and
$G^-$ given their particular embeddings in $\SO(d)$. One can follow
the detailed strategy of~\cite{Gauntlett:2002sc} to derive the
necessary and sufficient conditions on this $G$-structure in order
that the geometry preserves the corresponding supersymmetry. This is
based on direct manipulations of the Killing spinor equations and some
details of this approach appear in~\cite{Gauntlett:2001ur}. 

Equivalently, we can obtain the conditions on the $G$-structure
by writing the $G^\pm$-structures in terms of the $G$-structure and
then imposing the conditions on the $G^\pm$-structures derived in
section~\ref{sec:canon}.  In implementing this strategy it is crucial
to recall that the signs presented in section~\ref{sec:canon} assumed 
that the preserved spinors were of the $\e^+$-type
and also took, in the relevant cases, the preserved spinors to have a
definite chirality. In order to get the results of this section, one
needs the appropriate generalisations for $\nabla^-$ and sometimes 
the opposite chirality.

\smallsec{$\SU(2)$-geometries in $d=6$}
This case arises when both $\nabla^\pm$ have $\SU(3)$ holonomy with a
common $\SU(2)$ subgroup. The $\SU(2)$ structure in $d=6$ is specified
by a two-form $J$, a complex two-form $\Omega$ and two one-forms $K^i$
with $i=1,2$. They satisfy~\eqref{SUn} for $n=2$ and in addition  
\begin{equation}
\label{2SU3}
   i_{K^i} \Omega = i_{K^i} J = 0 .
\end{equation}
The corresponding $\SU(3)$ structures associated with $\nabla^\pm$ are
given by 
\begin{equation}
\begin{aligned}
   J^\pm &= J \pm K^1\wed K^2 , \\
   \Omega^\pm &= \Omega \wed (K^1 \pm iK^2) .
\end{aligned}
\end{equation}

Demanding that the $\SU(3)$-structures each satisfy the necessary and 
sufficient conditions for supersymmetry discussed in \p{SUncond}, \p{sunflux}
(with appropriate sign changes for $\nabla^-$, as mentioned above)
leads to necessary and sufficient conditions on the $SU(2)$ structure.
Specifically, we find
\begin{equation}
\label{2SU3cond}
\begin{aligned}
   \diff (\ex^{-\Phi}K^i ) & = 0 , \\
   \diff (\ex^{-\Phi}\Omega ) &= 0 , \\
   \diff J \wed K^1 \wed K^2 &= 0 .
\end{aligned}
\end{equation}
with the flux given by 
\begin{equation}
\label{2SU3H}
   *H = - \ex^{2\Phi}\diff \left(\ex^{-2\Phi} J \right) .
\end{equation}
These geometries
preserve two complex chiral $d=6$ spinors, one $\epsilon^+$ and
one $\epsilon^-$.

They also possess an  almost product structure
\begin{equation}
   \Pi = 2K^1\otimes K^1{}^\# + 2K^2\otimes K^2{}^\# - \id , 
\label{prodstr1}
\end{equation}
where $K^\#$ is the vector field dual to the one-form $K$, 
satisfying $\Pi\cdot\Pi=\id$. Since $\diff (\ex^{-\Phi}K^i )= 0$ this
structure is integrable and hence the metric can be cast in the
canonical form 
\begin{equation}
   \diff s^2 = g^4_{ab}(x,y)\,\diff x^a \,\diff x^b 
       + \ex^{2\Phi(x,y)} \delta_{ij}\dd y^i \dd y^j .
\label{fourtwometric} 
\end{equation}
The conditions \reef{2SU3cond} then imply that at fixed $y^i$,
the $SU(2)$ structure on the four-manifold has $W_2=W_4=0$ and
$W_5=\dd \Phi$. Such geometries, which in particular are \Ka,
are called almost Calabi--Yau.

This case corresponds to fivebranes wrapping \Ka-two-cycles in $CY_2$.
This is mirrored in the expression for the flux \p{2SU3H}, and
also in the structure of the metric \p{fourtwometric} with
the $y$ directions corresponding to the two directions transverse to
the fivebrane and the initial $CY_2$.
Explicit examples of such solutions were presented in 
\cite{Gauntlett:2001ps,zaffa} 
and were further explored from the world-sheet point of
view in \cite{Hori:2002cd}.

\smallsec{$SU(3)$-geometries in $d=7$}
This case arises when $\nabla^\pm$ each have $G_2$ holonomy and was
discussed in~\cite{Gauntlett:2002sc}. The $\SU(3)$ structure in $d=7$ is
specified by $J$ and $\Omega$ satisfying~\p{SUn} for $n=3$, and a vector
$K$ such that  
\begin{equation}
\label{2G2}
   i_K \Omega = i_K J = 0  .
\end{equation}
The two $G_2$ structures are given by
\begin{equation}
   \phi^\pm = J\wedge K\mp \im \Omega
\end{equation}
and demanding that they satisfy ~\p{G2cond}, \p{G2flux} (and their
generalisation for $\nabla^-$) 
leads to the differential conditions
\begin{equation}
\label{2G2cond}
\begin{aligned}
  \diff (\ex^{-\Phi} K) & = 0, \\
  \diff (\ex^{-\Phi} J) & = 0, \\
  \diff (\ex^{-\Phi}\re \Omega )\wed K & = 0,\\
  \diff (\im \Omega) \wed \im \Omega & = 0 
\end{aligned}
\end{equation}
with the flux given by 
\begin{equation}
\label{2G2H}
   *H = -\ex^{2\Phi} \diff (\ex^{-2\Phi} \im \Omega) .
\end{equation}
These geometries preserve
two $d=7$ spinors, one $\epsilon^+$ and one $\epsilon^-$.
The obvious almost product structure is again integrable 
and hence the metric can be cast in the canonical form
\begin{equation}
   \diff s^2 = g^6_{ab}(x,y)\,\diff x^a \,\diff x^b + \ex^{2\Phi(x,y)}\dd y^2~.
\label{sixonemetric} 
\end{equation}
The six-dimensional slices at fixed $y$
have an $SU(3)$ structure with intrinsic torsion lying in
${\cal W}_2\oplus {\cal W}_4\oplus {\cal W}_5$, and  
it is straightforward to see
that $W_4=-W_5=2\diff\Phi$. Recall that for $SU(3)$ the module
${\cal W}_2$ splits into two modules ${\cal W}_2^\pm$. The third
condition in \p{2G2cond} implies that while $W_2^+$ vanishes 
$W_2^-$ does not. These geometries are not Hermitian, as noted
in \cite{Gauntlett:2002sc}.
This case corresponds to fivebranes wrapping SLAG three-cycles and
explicit solutions were given in \cite{Gauntlett:2001ur,Gomis:2001aa}.

\smallsec{$SU(3)$-geometries in $d=8$}
This is one of the cases when $\nabla^\pm$ each have $SU(4)$
holonomy. It is in fact very similar to the case
of an $SU(2)$ structure in $d=6$ considered above.
The $SU(3)$ structure in $d=8$ is specified by $J,\Omega$ satisfying
\p{SUn} for $n=3$ and two vectors $K^i$ satisfying \p{2SU3}. 
The two $SU(4)$ structures are given by
\begin{equation}
\begin{aligned}
   J^\pm &= J \pm K^1\wed K^2 , \\
   \Omega^\pm &= \Omega \wed (K^1 \pm iK^2) .
\end{aligned}
\end{equation}
Demanding that they satisfy the necessary and sufficient conditions for $SU(4)$
structures given in \p{SUncond}, \p{sunflux} (and their generalisation
for $\nabla^-$) leads to the differential conditions as in \p{2SU3cond}
with the flux given by 
\begin{equation}
\label{2SU4H}
   *H = -\ex^{2\Phi}\diff \left(\ex^{-2\Phi}\tfrac{1}{2}J \wed J \right) . 
\end{equation}
Such geometries preserve two pairs of $d=8$ spinors with opposite
chirality, two $\epsilon^+$ and two $\epsilon^-$.
Again there is an integrable product structure and
the metric can be written in the form 
\begin{equation}
   \diff s^2 = g^6_{ab}(x,y)\,\diff x^a \,\diff x^b 
       + \ex^{2\Phi(x,y)} \delta_{ij}\dd y^i \dd y^j .
\label{sixtwometric} 
\end{equation}
At fixed $y^i$, the $SU(3)$ structure on the six-manifold is
almost Calabi--Yau, with the only non-vanishing class being $W_5=\dd \Phi$.
This case corresponds to fivebranes wrapping \Ka-four-cycles in $CY_3$
and solutions were found in \cite{Hori:2002cd,Naka:2002jz}.

\smallsec{$SU(2)\times SU(2)$-geometries in $d=8$}
The second way that $\nabla^\pm$ both have $\SU(4)$ holonomy is when they
give a common $SU(2)\times SU(2)$ structure. The
two orthogonal $SU(2)$ structures $J^A$ and
$J'{}^A$ satisfy the conditions~\eqref{2SU2struc}. 
The two $SU(4)$-structures are given by
\bea
J^\pm&=&J^3\pm J'{}^3\nn
\Omega^\pm&=&\Omega\wedge \Omega', \qquad \Omega\wedge \bar\Omega'
\eea
where e.g. $\Omega=J^2+\ii J^1$.
Demanding that they satisfy the necessary and sufficient conditions for $SU(4)$
structures given in \p{SUncond}, \p{sunflux} (and their generalisation
for $\nabla^-$) leads to the necessary and sufficient conditions on
the $SU(2)\times SU(2)$ structure 
given by
\begin{equation}
\label{su2su2conds}
\begin{aligned}
   \vol' \wed \diff J^3 & =  0 , \\
   \diff (\ex^{-\Phi}J^{A})  & =  0 && \qquad \text{for $A=1,2$}\\
   \vol \wed \diff J'{}^3 & =  0\\
   \diff (\ex^{-\Phi} J'{}^A)  & =  0 && \qquad \text{for $A=1,2$}
\end{aligned}
\end{equation}
where e.g. $\vol = \frac{1}{2}J^3\wedge J^3$.
These geometries preserve four
$d=8$ spinors with the same chirality, two $\e^+$ and two $\e^-$.
The almost product structure
\begin{equation}
   \Pi = J^+\cdot J^-=J^3 \cdot J^3 - J'{}^3 \cdot J'{}^3 ,
\label{prodstrgsu2su2}
\end{equation}
is integrable\footnote{Note that the existence of a generic pair
  $J^\pm$ of integrable complex structures satisfying only $[J^+,J^-]=0$ 
does not guarantee that the almost product structure $\Pi=J^+\cdot J^-$ is 
integrable. A concrete counter example is discussed in 
Appendix~\ref{app:product}.}
since $\nabla^\pm J^\pm=0$, $J^\pm$ commute and $J^\pm$ are integrable
(see Appendix~\ref{app:product}) and implies the canonical form of the metric 
\begin{equation}
   \diff s^2 = g^4_{ij}(x,y)\,\diff x^i \,\diff x^j 
      + {g'}^4_{ab}(x,y)\,\diff y^a \,\diff y^b ,
\label{susu2metric}
\end{equation}
each block being four-by-four. The four-dimensional slices each 
have an $SU(2)$ structure, with $W_2=W_4=0$ and $W_5=\dd \Phi$  
at any point in their transverse directions. 
The flux is given by 
\begin{equation}
\label{su2su2flux}
   *H = -\ex^{2\Phi} \diff (\ex^{-2\Phi} J^3 \wed J'{}^3) .
\end{equation}
These geometries arise when a fivebrane wraps a two-cycle in one
Calabi-Yau two-fold and a second two-cycle in a second Calabi-Yau two-fold.

\smallsec{$Sp(2)$-geometries in $d=8$}
This case arises when $\nabla^+$ has $SU(4)$ holonomy while
$\nabla^-$ has $\Spin(7)$ holonomy.
These correspond to fivebranes wrapping
C-LAG four-cycles in hyper-\Ka eight manifolds. Recall that these are
complex with respect to one complex structure and 
special Lagrangian with respect
to the remaining two. We have a $\Symp(2)$ structure given by a
triplet of complex structures satisfying~\eqref{Spstruc}.
The $SU(4)$ structure is given by $(J^3,\Omega^3)$, where
\begin{equation}
\Omega^3=\frac{1}{2}J^2\wedge J^2-\frac{1}{2}J^1\wedge J^1 +i(J^1\wedge J^2),
\end{equation}
and satisfies \p{SUncond}
while the \spin structure is defined by
\begin{equation}
\Psi=\frac{1}{2}J^1\wedge J^1+\frac{1}{2}J^2\wedge J^2 
-\frac{1}{2}J^3\wedge J^3
\end{equation}
which satisfies \p{spincond} (with appropriate sign changes).
This leads to 
the conditions on the $\Symp(2)$-structure given by
\begin{equation}
\label{spinSU4cond}
\begin{aligned}
   \diff \left( \ex^{-\Phi} J^A\right) &= 0 ,  \qquad \text{for $A=1,2$}\\
   \diff^\dagger (\ex^{-2\Phi} J^3) &= 0 ,
\end{aligned}
\end{equation}
and flux given by
\begin{equation}
\label{spinSU4H}
  *H= - \ex^{2\Phi}\diff \left(\ex^{-2\Phi}\re\Omega^1\right) 
       = \ex^{2\Phi}\diff \left(\ex^{-2\Phi}\re\Omega^2\right) 
       = -\ex^{2\Phi}\diff
          \left(\ex^{-2\Phi}\tfrac{1}{2}J^3\wed J^3\right) .
\end{equation}
These geometries preserve three $d=8$ spinors of the same chirality,
two $\epsilon^+$  and one $\epsilon^-$. Note that the conditions
imply that the two {\bf 16}'s in each of $\dd J^1$ and $\dd J^2$ vanish. 
Moreover, the six independent Lee forms are given by 
\begin{equation}
\begin{gathered}
   L^{11}=3\dd\Phi,\qquad L^{22}=3\dd\Phi,\qquad L^{33}=2\dd\Phi, \\
   L^{12}-L^{21}=-2J^3\cdot\dd\Phi,\qquad
   L^{31}-L^{13}=-J^2\cdot\dd\Phi,\qquad
   L^{23}-L^{32}=-J^1\cdot\dd\Phi.\qquad 
\end{gathered}
\end{equation}
It is worth emphasising that the intrinsic torsion of this
$Sp(2)$ structure is not totally anti-symmetric, and hence the geometry is not
HKT. It would be interesting to find explicit examples.

\smallsec{$SU(4)$-geometries in $d=8$}
This is the first case when $\nabla^\pm$ each have \spin holonomy. 
It corresponds to fivebranes wrapping
SLAG four-cycles in $CY_4$. In this case we have an $SU(4)$ structure
$J,\Omega$ satisfying
\p{SUn} for $n=4$.
The two \spin structures are given by
\begin{equation}
\Psi^\pm = \frac{1}{2}J\wedge J\pm \re\Omega
\end{equation}
and satisfy \p{spincond} (with sign changes for $\nabla^-$)
leading to the conditions on the $SU(4)$-structure
\begin{equation}
\label{2spinSU4cond}
\begin{aligned}
   \diff \left(\ex^{-\Phi}J \right) &= 0 , \\
   {} * ({} * \diff\re\Omega \wed \re\Omega ) &= - 6 \,\diff \Phi ,
\end{aligned}
\end{equation}
with flux given by
\begin{equation}
\label{2spinSU4H}
   * H = - \ex^{2\Phi}\diff \left(\ex^{-2\Phi}\re\Omega\right) .
\end{equation}
These geometries preserve two $d=8$ spinors with the same chirality,
one $\e^+$ and one $\e^-$. The intrinsic torsion of the $SU(4)$
structure lies in ${\cal W}_2 \oplus {\cal W}_4\oplus {\cal W}_5$, 
with $2W_4=W_5=6 \dd\Phi$, and so in particular the geometries are not
Hermitian.

\smallsec{$G_2$-geometries in $d=8$} 
This is the second case when $\nabla^\pm$ each have \spin holonomy. 
It occurs when fivebranes wrap co-associative four-cycles in
$G_2$ manifolds. In this case the two \spin structures give rise to
a $G_2$ structure with $\phi$ as in \reef{g2form} and
\begin{equation}
\label{2spinG2}
   i_K \phi = 0 .
\end{equation}
The two \spin structures are given by
\begin{equation}
\Psi^\pm=-i_K *\phi\pm\phi\wedge K
\end{equation}
and satisfy \p{spincond} (and sign changes for $\nabla^-$)
leading to the necessary and sufficient
conditions
\bea
\label{2spinG2cond}
   \diff (\ex^{-\Phi}K)  &=& 0 , \\
   \diff (\ex^{-\Phi}\phi) \wed K &=& 0 , \\
   {} * \diff (i_K * \phi) \wed \phi \wed  K &=& 0 , \\
   {} * \diff (i_K * \phi) \wed i_K * \phi &=& 4 * \diff \Phi ,
\eea
with flux given by
\begin{equation}
\label{2spinG2H}
   * H = \ex^{2\Phi}\diff \left(\ex^{-2\Phi} i_K * \phi\right) .
\end{equation}
These geometries preserve one $\e^+$ and one
$\e^-$ $d=8$ spinor of opposite chirality.
The intrinsic torsion of the $G_2$ structure lies in ${\cal W}_2\oplus
{\cal W}_4$ with $W_4=-4 \dd \Phi$. This means one cannot introduce a
$G_2$ Dolbeault cohomology~\cite{fernug}.

\smallsec{$\{\id\}$-geometries} 
For completeness let us briefly mention the case corresponding to
the first entry in table~\ref{blue}. This case has 
two different $\SU(2)$ structures each satisfying
\p{SUn} giving a trivial structure defined by four real $K^i$ vectors. A
little calculation reveals that this case can always be put in the
canonical form 
\begin{equation}
\begin{aligned}
\label{flatNSmetric}
   \diff s^2 &= \ex^{2\Phi}\diff s^2 (\bbR^{4})~, \\
   * H &= - \ex^{2\Phi} \diff (\ex^{-2\Phi}) ,
\end{aligned}
\end{equation}
which is just the transverse space to the simple fivebrane solution.

\medskip

We conclude this section with two comments. First, considering either
set of $\epsilon^+$ or $\epsilon^-$ Killing spinors we see that the
geometries of this section are special cases of those appearing in
section~\ref{sec:canon}. It is then clear, from the results of
section~\ref{sec:gen}, that supersymmetric geometries in $d=9$ can be
obtained by fibering an appropriate number of flat directions over the
geometries in this section. In order that the same amount of
supersymmetry is preserved, the fibrations are determined by Abelian
instantons that satisfy the generalised self-duality conditions for
both of the $G^\pm$-structures. In other words they must satisfy the 
generalised self-duality conditions for the maximal common subgroup
$G$. Note that in general the Bianchi identity for $H$ may further
restrict which fibrations are possible. For instance in the cases
where both $\nabla^+$ and $\nabla^-$ have $\SU(n+1)$ holonomy, one can
show that $\dd H$ has no components transforming as a four-form under
$\SO(2n)\supset\SU(n)$ for the common $\SU(n)$-structure. As such,
there are in fact no solutions with non-trivial twisting.

The second comment is to note that we have only considered structures
$G^\pm$ that are orthogonal in the sense that preserved spinors $\e^+$
and $\e^-$ are orthogonal, that is $\bar{\epsilon}^+\epsilon^-=0$. In
fact, as we now show, this is a necessary condition for a non-trivial
solution to be supersymmetric. Take any two Killing spinors
$\epsilon^+$ and $\epsilon^-$. The vanishing of the gravitini
variations implies that 
\begin{equation}
   \nabla_m (\bar{\epsilon}^+\epsilon^-) 
      = \frac{1}{4}H_{mab}\bar\e^+\gamma^{ab}\e^- .
\end{equation}
The dilatino equation implies that for any gamma-matrix operator $A$
we have 
\begin{equation}
   \partial_m\Phi {\bar\epsilon^+}[A,\gamma^m]_\pm\epsilon^-
      =\frac{1}{12}H_{mnp}
          {\bar\epsilon^+}[A,\gamma^{mnp}]_\pm\epsilon^- .
\end{equation}
Taking $A=\gamma^m$ and using the upper sign, we conclude that
\begin{equation}
   \nabla_m (\bar{\epsilon}^+\epsilon^-) 
      =\partial_m\Phi (\bar{\epsilon}^+\epsilon^-) .
\end{equation}
This is trivially satisfied if the $G^\pm$-structures are orthogonal
since then $\bar{\epsilon}^+\epsilon^-=0$. If the structures are not
orthogonal, we have some point where $\bar{\epsilon}^+\epsilon^-$ is
non-zero and then by continuity there will be a neighbourhood in which
it is non-zero. In this neighbourhood we have
$\bar{\epsilon}^+\epsilon^-=\ex^{\Phi+\Phi_0}$, for some constant
$\Phi_0$. 

The two spinors $\epsilon^\pm$ define a pair of $G^\pm$-structures
both of which are sub-bundles of the same $\SO(d)$-bundle of
orthonormal frames  defined by the metric $g_d$\footnote{Note that this
  is only true for $D^\pm\epsilon^\pm=0$ with $D^\pm$ a pair of
  spin-connections, compatible with the metric $g_d$, and not, for
  instance, if $D^\pm$ are general Clifford connections.}. 
Together $\epsilon^\pm$ define a common $G$-structure sub-bundle of
the two $G^\pm$-structures. Furthermore, there always exists
\emph{some} metric-compatible connection $\tilde\nabla$ that preserves
this $G$-structure. (Note this connection generically does not have
totally antisymmetric torsion.) Necessarily it preserves the
$G^\pm$-structures, so that $\tilde\nabla\e^\pm=0$. Thus in fact we have
$\nabla(\bar{\e}^+\e^-)=\tilde\nabla(\bar\e^+\e^-)=0$ implying $\Phi$
is a constant. However, the equations of motion then imply that 
$H$ is constant.  We thus conclude that there are no supersymmetric
solutions with non-vanishing flux when the structures $G^\pm$ are not
orthogonal.


\section{Explicit examples II}
\label{sec:ex2}


In this section, we present some further explicit solutions in $d=6$,
some preserving both $\e^+$ and $\e^-$ supersymmetries, for the type II
theories, including a solution that preserves the unusual fraction of
${12/32}$ supersymmetry. The basic solutions have two flat directions
fibred over a four dimensional base-space, with the fibration
being specified by two Abelian instantons on the base, and thus
generalise those discussed in section~\ref{sec:ex1}. 
We shall also discuss compact heterotic geometries 
in $d=6$ preserving both eight and four supercharges. 

It will be convenient in this section to distinguish different
six-dimensional solutions by the number of preserved
supersymmetries. Let us start with the most supersymmetric case
corresponding to a flat NS five-brane as discussed at the end of
the last section. Recall that the $d=4$ solution transverse to a
simple fivebrane~\p{flatNSmetric} preserves eight $\e^+$ spinors and
eight $\e^-$ spinors satisfying the projections
\begin{equation}
   \gamma^{1234}\e^+=-\e^+,\qquad \gamma^{1234}\e^-=+\e^- .
\end{equation}
As previously noted $\nabla^\pm$ have $\SU(2)^{\pm}$ holonomy in
$\SO(4)=\SU(2)^+\times\SU(2)^-$ with the maximal common subgroup
being the identity. We can trivially lift this 
to a six-dimensional solution by adding two extra flat
directions. This still preserves 16 supercharges corresponding to ${\cal N}=4$
supersymmetry in the remaining four spacetime dimensions.

We now twist the two flat directions, as in section~\ref{sec:ex1}, with
two Abelian instantons, 
\begin{equation}
\label{n=3} 
\begin{aligned}
   \dd s^2 &= \ex^{2\Phi}\dd \tilde s^2 
      +(\dd y+B^1)^2 + (\dd z+B^2)^2 , \\
   H_{mnp} &= -\tilde\epsilon_{mnp}{}^q\tilde{\nabla}_q \ex^{2\Phi}
      + 3B^1_{[m}G^1_{np]} + 3B^2_{[m}G^2_{np]} , \\
   H_{mny} &= G^1_{mn} , \qquad \qquad
   H_{mnz} = G^2_{mn} ,
\end{aligned}
\end{equation}
giving the dilaton equation
\begin{equation}
   \tilde{\nabla}^2\ex^{2\Phi} = 
      - \frac{1}{2}\left((\tilde{G}^1)^2 +(\tilde{G}^2)^2\right) ,  
\end{equation}
where $m,n=1,\dots 4$ and now $G^i=\dd B^i$ are taken to be
\textit{self-dual} instantons on the $\bbR^4$ base space
$\dd\tilde{s}^2$. This twisting still preserves eight $\e^-$ spinors
so that $\nabla^-$ still has $\SU(2)^-$ holonomy. For non-zero $G^i$,
generically the  solution however breaks all of the $\e^+$
supersymmetry. (Note, simply for convenience of later discussion, 
we have exchanged the roles of $\nabla^+$ and
$\nabla^-$, by taking $H\to -H$ and changing the orientation on the base,
as compared to the discussion in
section~\ref{sec:ex1}. There we took anti-self-dual instantons so that
$\epsilon^+$ spinors were preserved. This accounts for the difference
in signs of the terms involving $B$ and $G$ in~\p{n=3} compared to
those in~\p{explictd=5ex}). Hence, generically these solutions
preserve ${\cal N}=2$ supersymmetry in the remaining four spacetime
dimensions. 

Interestingly, it is nonetheless possible to preserve four $\e^+$
Killing spinors corresponding to $\nabla^+$ having $\SU(3)$ holonomy,
for suitably chosen non-generic instantons. To see this we define an
$\SU(3)$ structure 
by
\begin{equation}
\begin{aligned}
   J &= \ex^{2\Phi}\tilde{J} + (\dd y+B^1) \wedge (\dd z+B^2) , \\
   \Omega &= \ex^{2\Phi}\tilde{\Omega} \wedge 
      \left[(\dd y+B^1)+ \ii(\dd z+B^2) \right] ,
\end{aligned}
\end{equation}
where $\tilde{J}=\dd x^1\wedge\dd x^2+\dd x^3\wedge\dd x^4$ and
$\tilde\Omega=(\dd x^1+\ii\dd x^2)\wedge (\dd x^3+\ii \dd x^4)$ define the
$\SU(2)^+$ structure on $\bbR^4$. Demanding that the $\SU(3)$ structure
satisfies the conditions for supersymmetry~\p{SUncond}, we find that
\begin{equation}
\begin{aligned}
   \tilde{J} \lrcorner\; G^i &= 0 , \\
   \tilde{\Omega}\lrcorner\; (G^1+\ii G^2) &= 0 .
\end{aligned}
\end{equation}
The generic constant flux solution to these equations is given by
\begin{equation}
\label{spchoice}
   G^1 + \ii G^2 = k \tilde{\Omega}
\end{equation}
for some complex constant $k$. (Note, as we discuss below, this is the
same twisting that appears in the Iwasawa manifold analysed
in~\cite{dallagata}.)  The Bianchi identity then implies the equation
for the dilaton 
\begin{equation}
\label{dileqhere}
   \tilde{\nabla}^2\ex^{2\Phi} = -8|k|^2 ,
\end{equation}
which can easily be solved. To summarise, the solution~\p{n=3} with flat base
space will
preserve eight $\epsilon^-$ and four $\epsilon^+$ spinors for the
specific choice of self-dual instantons \p{spchoice} and dilaton
satisfying~\p{dileqhere}. 

A number of comments are now in order. First, 
this special solution corresponds to ${\cal N}=3$ supersymmetry in the
remaining four spacetime dimensions. It would be interesting to relate
this solution to those discussed in~\cite{Frey:2002hf}. 

Secondly, the holonomy of the connections
$\nabla^\pm$ for the special solution are $\SU(3)$ and $\SU(2)$,
respectively. This is not a combination appearing in
table~\ref{blue}. The form of the solution indicates that this
solution is related to fivebranes wrapping two flat directions, but a
world-volume interpretation of the twisting and preservation of
supersymmetry are obscure to us at present. 

Thirdly, this special background is also a
heterotic/type~I solution. In this case, one loses the $\epsilon^-$
supersymmetries and the solution preserves only four $\epsilon^+$
spinors, and so has ${\cal N}=1$ supersymmetry in four-dimensions. Including
additional heterotic instantons simply add to the source $|k|^2$ in
the dilaton equation~\eqref{dileqhere}. Note that by taking $H\to -H$ and
switching the orientation of the base, we
switch $\e^+$ and $\e^-$ and hence we can also
obtain a heterotic solution from the generic solution
\p{n=3} with an $SU(2)$ structure and
${\cal N}=2$ superymmetry.

Finally, the metric and three-form obtained by setting the dilaton to
constant in~\eqref{n=3} with $G^1+\ii G^2=k\tilde{\Omega}$, were
first considered in the heterotic case (including an additional Abelian
instanton embedded in $E_8\times E_8$ or $\SO(32)$)
in~\cite{dallagata}. There it was demonstrated that the conditions for
the preservation of $\e^+$ supersymmetry with $\nabla^+$ having
$\SU(3)$ holonomy were satisfied. However, given the analysis here,
the background in \cite{dallagata} is problematic for the following somewhat
subtle reason. As we have already noted when the dilaton is constant and
$H\neq0$, the leading-order type~II (or heterotic/type~I) equations of
motion are not satisfied. As shown in~\cite{Gauntlett:2002sc}, these
equations of motion are a direct consequence of the preservation of
supersymmetry once the Bianchi identity~\eqref{eq:BI1} is imposed (or
equivalently~\eqref{eq:BI2} if  $\tr R\wedge R=0$ as for the geometry 
considered in~\cite{dallagata}). This contradiction is resolved by the
fact that the background in~\cite{dallagata} actually satisfies a
Bianchi identity with the opposite sign to the one arising in type I
supergravity. This discrepancy is probably related to the sign
discrepancy between the expression~\eqref{strom} and the corresponding
expression in~\cite{strominger}\footnote{Following recent
  correspondence the authors of~\cite{dallagata} have independently
  confirmed this discrepancy in~\cite{strominger}.}.

The type II solutions we have been discussing
can also be generalised by replacing the
flat space in~\p{n=3} with a generic Calabi--Yau two-fold $CY_2$. 
As usual for type II, the Calabi-Yau two-fold cannot be compact
in order to satisfy the Bianchi identity $\dd H=0$. If
we take the orientation of the $CY_2$ to be such that the complex
structures are self-dual, we impose the projections
$\gamma^{1234}\epsilon^\pm=-\e^\pm$. In this case, the solution
preserves no  $\e^-$ supersymmetry, and generically no $\e^+$
supersymmetry. However, choosing 
$G^1+\ii G^2=k\tilde{\Omega}$, where $\tilde{\Omega}$ is the
holomorphic $(2,0)$ form on $CY_2$, we find that $\nabla^+$ has
$\SU(3)$ holonomy and the solution still preserves four $\e^+$
supersymmetries, corresponding to ${\cal N}=1$ supersymmetry in four
dimensions.  Alternatively, if the orientation of the $CY_2$ is chosen
so that the complex structures are anti-self-dual, we impose the
projections $\gamma^{1234}\epsilon^\pm=+\e^\pm$. These solutions break
all of the $\e^+$ supersymmetry, but preserve eight $\e^-$ spinors. The
latter choice of orientation corresponds (after exchanging $\e^+$ with
$\e^-$ by taking $H\to -H$ and switching the orientation on the base) 
to a simple generalisation from $d=5$ to
$d=6$ of the solutions discussed in section~\ref{sec:ex1} and
explicitly obtained 
in~\cite{Lu:2002xa} for the cases of Taub--NUT and Eguchi--Hanson. The
former choice of orientation on the other hand, gives a new kind of
supersymmetric solution that exploits the fact that one is twisting
two flat directions and not just one as considered
in~\cite{Lu:2002xa}.

Similarly, one can obtain heterotic/type~I geometries preserving
${\cal N}=1,2$ supersymmetry. By taking the flat directions to be a
two-torus, and $M_0$ to to be either conformally $T^4$ or conformally
$K3$, we get compact and supersymmetric heterotic geometries. It will
be interesting to see whether it is possible to solve the heterotic
Bianchi identity for these geometries; if it is, as in
section~\ref{sec:ex1}, the $\tr R\wedge R$ contribution will be
essential. In addition, one should again find that the radius of the
two-torus is required to be of order the string scale and that several
of the moduli are fixed.

\section{Discussion}
\label{sec:concl}

In this paper we have studied the necessary and sufficient conditions
for static geometries of type I/heterotic string theory, or type
II theories with only non-vanishing NS-NS fields, to preserve supersymmetry
and solve the equations of motion. The Killing spinors define 
$G$-structures on the geometries
and we determined the intrinsic torsion of the $G$-structure. 
We emphasised the universal expression for the three-form flux in 
terms of generalised calibrations and the connection with
wrapped branes, following \cite{Gauntlett:2001ur,Gauntlett:2002sc}.
This universal expression for the flux leads to
a very simple proof of a vanishing theorem on compact 
manifolds.

The geometries always have a connection with totally anti-symmetric torsion, 
$\nabla^+$ (or $\nabla^-$ for the type II theories),
which has special holonomy. We first discussed the geometries
in the canonical dimension for the special holonomy group, $d=8$
for \spin, $d=7$ for $G_2$, etc. We then showed that the most general
geometries in $d=9$ have a number of flat directions fibred over these
geometries in the canonical dimensions, with the fibration being determined
by Abelian generalised instantons.
We also discussed the
physical interpretation of these geometries in terms of wrapped
fivebranes. For example, the eight-dimensional
geometries with a single flat dimension 
fibred over a seven-dimensional geometry with $G_2$-structure
correspond to fivebranes wrapping supersymmetric cycles of the 
form $S^1\times \Sigma_3\subset S^1\times M_{G_2}$ where $\Sigma_3\subset
M_{G_2}$ is an associative three-cycle in a $G_2$-holonomy manifold. The fact
that the resulting eight-dimensional
geometry is not necessarily a direct product of $S^1$ with a seven-dimensional
geometry is worth further investigation.
We presented some explicit examples, that would be worth studying
further and generalising.

These results provide a comprehensive classification
of all of the supersymmetric static geometries
of the heterotic/type I theory. For the type II theories, 
we also analysed the geometries that arise when both
connections $\nabla^\pm$ have special holonomy. Our analysis
covers all cases of NS fivebranes wrapping calibrated cycles, as listed
in tables \ref{green} and \ref{blue}. 

We also presented an explicit solution with a torus $T^2$ fibred over
an $\bbR^4$ base with $\nabla^+$ having $\SU(3)$ holonomy and
$\nabla^-$ having $\SU(2)$ holonomy. This solution has 
four $\e^+$ Killing spinors and eight $\e^-$ spinors. The form of the
flux suggests that the solution should be interpreted as a flat fivebrane
with two of the world-volume directions further wrapped on the
two torus. Naively, one would therefore expect 8 plus 8  Killing spinors
and so it would also be interesting to find a physical interpretation of the
twisting which leads to this reduction of supersymmetry. In
\cite{Frey:2002hf} type II solutions on $T^6$ orientifolds with
non-vanishing R-R and NS-NS fluxes were presented that also preserve
12 Killing spinors and it would be interesting to see if they are
related. Perhaps our solutions provide a local descriptions of blow-ups
of geometries around certain fixed points. 

Candidate heterotic compactifications in $d=6$ were also presented,
preserving both four and eight supersymmetries. They are based on
manifolds which are fibrations of $T^2$ over a $K3$ base. The models
with four supersymmetries arise for non-generic complex structure on
the $K3$ and there are additional constraints on the radii of the
circles of the torus. This indicates that many moduli are fixed. We
showed that the size of the torus is necessarily of order the string
scale, indicating that the supergravity approximation is breaking
down. One would also has to check the equations for motion are satisfied. To
pursue these models further we aim to construct a conformal
field theory description. It would also be interesting to relate our
compactifications to those of~\cite{BeckDas,trivedi,becker}. 

We have emphasised that the expression for the three-form flux
is easy to understand as a generalised calibration since the geometry
should still admit fivebranes wrapping the corresponding cycles. It is
very interesting to note that many, and in some cases all, of the other
conditions constraining the intrinsic torsion can be interpreted in
the same way. For example, consider the case of the $\SU(3)$ structure
with only $\epsilon^+$ Killing spinors. The expression for the
flux~\eqref{sunflux} is the general calibration condition for a
fivebrane wrapping a K\"ahler two-cycle in a 
Calabi--Yau three-fold. In addition the intrinsic torsion is constrained
to satisfy~\eqref{SUncond}. Suppose we consider the trivial product of
our $\SU(3)$ manifold $M_6$ with a torus $T^2$. Let $K^1=\dd y^1$
and $K^2=\dd y^2$ represent the extra directions. The full set of
conditions on the structure can then be written on the
eight-dimensional space $M_6\times T^2$ as  
\begin{equation}
\begin{aligned}
   \dd [\ex^{-2\Phi}J\wedge J] &= 0 , \\ 
   \dd [\ex^{-2\Phi}\Omega \wedge (K^1+\ii K^2)] &= 0 , \\
   \dd [\ex^{-2\Phi}J\wedge K^1 \wedge K^2] &= - \ex^{-2\Phi} * H . \\ 
\end{aligned}
\end{equation}
Given that $H$ lies solely in $M_6$, we see that all three expressions
are calibration conditions of the form
$*H=\ex^{2\Phi}\dd(\ex^{-2\Phi}\Xi)$ just for wrapping different
cycles. The first is for a fivebrane wrapping a K\"ahler four-cycle in
the Calabi--Yau, the second for wrapping a special Lagrangian cycle
(and one of the $K^i$ directions), while the last is the familiar
expression for the wrapping of a K\"ahler two-cycle in the Calabi--Yau
together with the torus $T^2$. This is physically reasonable, since
the geometry $M_6\times T^2$, 
corresponding to the full backreaction solution around a
brane wrapping a K\"ahler two-cycle, should still admit probe branes
wrapping the special Lagrangian three- and K\"ahler
four-cycles. Similar arguments extend to the fibration cases in
section~\ref{sec:gen} and the geometries with $\epsilon^+$ and
$\epsilon^-$ in section~\ref{sec:+-}. 

An important motivation for this work is that a good understanding of
the geometry underlying supergravity configurations might allow
us to find new explicit solutions. Indeed for the cases listed in 
table~\ref{green}
a co-homogeneity one ansatz is useful for finding solutions 
\cite{Gauntlett:2002sc}. This is a practical alternative to finding
solutions describing wrapped fivebranes
using the gauge supergravity approach initiated in \cite{Maldacena:2000mw}.
For the cases in table~\ref{blue}, on the other hand, 
a simple generalisation of this technique can lead to co-homogeneity one
but also to a co-homogeneity two or more ansatz, and progress in
the latter case is much more difficult \cite{Gauntlett:2002sc}. 
At present the gauge supergravity approach
is the best available tool to produce solutions for these latter cases. 
It should be noted, however, that since the configurations in 
table~\ref{blue} preserve more supersymmetry than those in table~\ref{green},
one expects that with new techniques, ultimately, they could 
be easier to analyse. 

Finally, it is natural to generalise this work to also include RR fields
in the type II theories, as well as to consider Lorentzian
geometries. Such geometries will allow one to describe both wrapped NS
and D-branes, as well pp-waves and general non-static backgrounds. 
Based on this work and on 
\cite{jerstas} we expect generalised calibrations to play an important role.

\subsection*{Acknowledgments}
We would like to thank Atish Dabholkar, 
Jan Gutowski and James Sparks for useful discussions.
D.~M.~ is supported by an EC Marie Curie Individual Fellowship under contract 
number HPMF-CT-2002-01539. D.~W.~is supported by a Royal Society
University Research Fellowship.


\appendix

\section{Equations of motion}
\label{app:conv}

The low-energy effective action for heterotic/type I string theory is
given by the type I supergravity action
\bea\label{acthere}
S=\frac{1}{2\kappa^2}\int d^{10} x \sqrt {-g}
e^{-2\Phi} \left(R+4(\nabla\Phi)^2-\tfrac{1}{12}H^2 - \alpha' \Tr
F^2\right) 
\eea 
where $F$ is in the adjoint of $SO(32)$ or $E_8\times E_8$. 
In type I supergravity the three-form $H$ satisfies a modified
Bianchi identity
\bea 
\dd H=2\alpha' \Tr F\wedge F .
\eea
Including the leading order string correction from anomaly cancellation
we get
\bea 
\dd H=2\alpha' (\Tr F\wedge F -\tr R\wedge R)
\eea
but to fully consistently implement this one should
also include modifications to the action. 
The equations of motion coming from \p{acthere} 
are given by 
\begin{subequations}
\begin{align}
   R_{MN} - \tfrac{1}{4}H_{MRS}H_{N}{}^{RS} + 2\nabla_M\nabla_N\Phi 
      -2 \alpha' \Tr F_M{}^R F_{NR} 
      &= 0 , \\
   \nabla^2(\ex^{-2\Phi}) - \tfrac{1}{6}\ex^{-2\Phi}H_{MNR}H^{MNR}
      - \alpha'\ex^{-2\Phi}\Tr F_{MN}F^{MN}
      &= 0 , \label{Phieom} \\
   \nabla_M (\ex^{-2\Phi}H^{MNR}) &= 0 , \\
   D^M (\ex^{-2\Phi}F_{MN}) - \tfrac{1}{2}\ex^{-2\Phi}F^{RS}H_{RSN}
      &=0 .
\end{align}
\end{subequations}
The action and equations of motion for the type II theories with
all RR fields set to zero are obtained by simply setting the gauge
field $F$ to zero and using the Bianchi identity $\dd H=0$.


\section{Spinor and $G$-structure conventions}
\label{app:spinor}


In doing calculations it is often useful to have an explicit set of
projections defining the Killing spinors and the corresponding
$G$-structures. Here we define one possible set of conventions
consistent with the expressions given in the paper. In particular, we
will use the same set of projectors (or subset of them) to define the
invariant spinors in all cases. Specifically, the Killing spinors will
be defined by their $\pm1$ eigenvalues for the set of commuting gamma
matrices  
\begin{equation}
   \gamma^{1234}, \gamma^{5678}, \gamma^{1256}, \gamma^{1357} . 
\end{equation}
We concentrate on the cases of $G$-structure in canonical
dimension. However, in each case we also give how the structure embeds
in the next simplest structure group following figure~\ref{fig}. Using
these embeddings one can obtain conventions for any of the
$G$-structures in arbitrary dimensions $d\leq 9$. 

Note that in all dimensions the gamma matrix algebra is taken to be
$\{\gamma_m,\gamma_n\}=2\delta_{mn}$ and the adjoint spinor is written
as $\bar{\epsilon}$ and the conjugate spinor as $\epsilon^\conj$. We
always normalise the Killing spinors to satisfy
$\bar{\epsilon}\epsilon=1$. 

\smallsec{$\Spin(7)$} 
In eight dimensions, a $\Spin(7)$-structure defines a single real
chiral invariant spinor $\epsilon$. For definiteness, we choose
$\gamma_{1\cdots8}\epsilon=\epsilon$. A possible set of independent, 
commuting projections defining a $\epsilon$ are 
\begin{equation}
   \gamma^{1234}\epsilon = \gamma^{5678}\epsilon =
   \gamma^{1256}\epsilon = \gamma^{1357}\epsilon = - \epsilon .
\end{equation}
Writing the Cayley four-form $\Psi$ as
\begin{equation}
\label{spin7s}
   \Psi_{mnpq} = - \bar{\epsilon}\gamma_{mnpq}\epsilon 
\end{equation}
then matches the expression~\eqref{Psidef}. The corresponding volume
form is given by 
\begin{equation}
\label{d8vol}
   \vol_{m_1\dots m_8} = \bar{\epsilon}\gamma_{m_1\dots m_8}\epsilon ,
\end{equation}
Note one can always choose a real basis for the gamma matrices so that
$\bar{\epsilon}=\epsilon^\trsp$. The conventions for lifting a
$\Spin(7)$-structure to $d=9$ are given in section~\ref{sec:d=9}.

\smallsec{$\SU(4)$}
An $\SU(4)$-structure leaves invariant two real orthogonal spinors
$\epsilon_{(a)}$ with $a=1,2$ of the same chirality in $d=8$. These
can be defined by  
\begin{equation}
   \gamma^{1234}\epsilon_{(a)} = \gamma^{5678}\epsilon_{(a)} =
   \gamma^{1256}\epsilon_{(a)} = - \epsilon_{(a)} 
\end{equation}
with 
\begin{equation}
   \gamma^{1357}\epsilon_{(1)} = + \epsilon_{(1)} , \qquad
   \gamma^{1357}\epsilon_{(2)} = - \epsilon_{(2)} .
\end{equation}
Defining a complex spinor
$\eta=\frac{1}{\sqrt{2}}(\epsilon_{(1)}+\ii\epsilon_{(2)})$, the forms
$J$ and $\Omega$ can then be written as 
\begin{equation}
\begin{aligned}
   J_{mn} &= -\ii\bar{\eta}\gamma_{mn}\eta , \\
   \Omega_{mnpq} &= \bar{\eta}^\conj\gamma_{mnpq}\eta .
\end{aligned}
\end{equation}
Note in the basis where $\bar{\epsilon}=\epsilon^\trsp$, we have the
more  familiar expressions $J_{mn}=\ii\eta^\dag\gamma_{mn}\eta$ and
$\Omega_{mnpq}=\eta^\trsp\gamma_{mnpq}\eta$. 
Given $\gamma^{12}\epsilon_{(1)}=-\epsilon_{(2)}$ we get the standard
expressions 
\begin{equation}
\begin{aligned}
   J &= e^{12} + e^{34} + e^{56} + e^{78} , \\
   \Omega &= (e^1+\ii e^2)(e^3+\ii e^4)(e^5+\ii e^6)(e^7+\ii e^8) .
\end{aligned}
\end{equation}
The corresponding volume form is given by~\eqref{d8vol} as above. Note
that each real spinor $\epsilon_{(a)}$ also defines a corresponding 
$\Spin(7)$-structure as in~\eqref{spin7s} given by  
\begin{equation}
\begin{aligned}
   \Psi^{(1)} = \tfrac{1}{2}J \wedge J - \re \Omega , \\   
   \Psi^{(2)} = \tfrac{1}{2}J \wedge J + \re \Omega .
\end{aligned}
\end{equation}

\smallsec{$\Symp(2)$}
We now have three real orthogonal invariant spinors $\epsilon_{(a)}$
with $a=1,2,3$ of the same chirality in $d=8$. These can be defined by  
\begin{equation}
   \gamma^{1234}\epsilon_{(a)} = \gamma^{5678}\epsilon_{(a)} 
      = \left(\gamma^{1256}
         +\gamma^{1357}+\gamma^{1458}\right)\epsilon_{(a)} 
      = - \epsilon_{(a)} 
\end{equation}
with 
\begin{equation}
   \gamma^{1256}\epsilon_{(1)} = + \epsilon_{(1)} , \qquad
   \gamma^{1458}\epsilon_{(2)} = + \epsilon_{(2)} , \qquad
   \gamma^{1357}\epsilon_{(3)} = + \epsilon_{(3)} .
\end{equation}
Note, the eigenvalues under $(\gamma^{1256},\gamma^{1357},\gamma^{1458})$ of
$\epsilon_{(a)}$ are $(+1,-1,-1)$, $(-1,-1,+1)$ and $(-1,+1,-1)$ for $a=1,2,3$
respectively. The three two-forms $J^A$ are then given by 
\begin{equation}
\begin{aligned}
   J^1_{mn} &= - \bar\epsilon_{(2)}\gamma_{mn}\epsilon_{(3)} , \\
   J^2_{mn} &= - \bar\epsilon_{(3)}\gamma_{mn}\epsilon_{(1)} , \\
   J^3_{mn} &= - \bar\epsilon_{(1)}\gamma_{mn}\epsilon_{(2)} .
\end{aligned}
\end{equation}
Given
$\gamma^{12}\epsilon_{(2)}=\gamma^{56}\epsilon_{(2)}=\epsilon_{(3)}$, 
$\gamma^{14}\epsilon_{(3)}=\gamma^{58}\epsilon_{(3)}=\epsilon_{(1)}$,
and
$\gamma^{13}\epsilon_{(1)}=\gamma^{57}\epsilon_{(1)}=\epsilon_{(2)}$,  
we have the explicit expressions
\begin{equation}
\begin{aligned}
   J^1 &= e^{12} + e^{34} + e^{56} + e^{78} , \\
   J^2 &= e^{14} + e^{23} + e^{58} + e^{67} , \\
   J^3 &= e^{13} + e^{42} + e^{57} + e^{86} .
\end{aligned}
\end{equation}
The corresponding volume form is given by~\eqref{d8vol} as above. 
Note that each almost complex structure $J^A$ as a $\SU(4)$-structure
has a corresponding $(4,0)$-form $\Omega^A$ given by 
\begin{equation}
\begin{aligned}
   \Omega^1 &= \tfrac{1}{2}J^3\wedge J^3 - \tfrac{1}{2}J^2\wedge J^2
      + \ii J^2\wedge J^3 , \\
   \Omega^2 &= \tfrac{1}{2}J^1\wedge J^1 - \tfrac{1}{2}J^3\wedge J^3
      + \ii J^3\wedge J^1 , \\
   \Omega^3 &= \tfrac{1}{2}J^2\wedge J^2 - \tfrac{1}{2}J^1\wedge J^1
      + \ii J^1\wedge J^2 .
\end{aligned}
\end{equation}
Each spinor $\epsilon_{(a)}$ also defines a corresponding
$\Spin(7)$-structure given by 
\begin{equation}
\begin{aligned}
   \Psi^{(1)} &= \tfrac{1}{2}J^2\wedge J^2 + \tfrac{1}{2}J^3\wedge J^3 
      - \tfrac{1}{2}J^1\wedge J^1 , \\
   \Psi^{(2)} &= \tfrac{1}{2}J^3\wedge J^3 + \tfrac{1}{2}J^1\wedge J^1  
      - \tfrac{1}{2}J^2\wedge J^2 , \\
   \Psi^{(3)} &= \tfrac{1}{2}J^1\wedge J^1 + \tfrac{1}{2}J^2\wedge J^2 
      - \tfrac{1}{2}J^3\wedge J^3 .
\end{aligned}
\end{equation}

\smallsec{$\SU(2)\times\SU(2)$}
We now have four orthogonal, real invariant spinors all of the same
chirality in $d=8$. They can be defined by 
\begin{equation}
   \gamma^{1234}\epsilon_{(a)} = \gamma^{5678}\epsilon_{(a)} 
      = - \epsilon_{(a)} 
\end{equation}
with 
\begin{equation}
   \gamma^{1256}\epsilon_{(a)} = \begin{cases}
      - \epsilon_{(a)} & \text{for $a=2,3$} \\
      + \epsilon_{(a)} & \text{for $a=1,4$} 
      \end{cases} ,
   \qquad
   \gamma^{1357}\epsilon_{(a)} = \begin{cases}
      - \epsilon_{(a)} & \text{for $a=1,2$} \\
      + \epsilon_{(a)} & \text{for $a=3,4$} 
      \end{cases} .
\end{equation}
The three two-forms $J^A$ are given by combinations, self-dual on
the $(a)$ index,
\begin{equation}
\begin{aligned}
   J^1_{mn} &=  - \left( \bar\epsilon_{(2)}\gamma_{mn}\epsilon_{(3)}
      + \bar\epsilon_{(1)}\gamma_{mn}\epsilon_{(4)} \right) , \\
   J^2_{mn} &= -\left(\bar\epsilon_{(3)}\gamma_{mn}\epsilon_{(1)}
      + \bar\epsilon_{(2)}\gamma_{mn}\epsilon_{(4)} \right) , \\
   J^3_{mn} &= - \left(\bar\epsilon_{(1)}\gamma_{mn}\epsilon_{(2)}
      + \bar\epsilon_{(3)}\gamma_{mn}\epsilon_{(4)} \right) . 
\end{aligned}
\end{equation}
The second set of $J'{}^A$ two-forms are given by the corresponding
anti-self-dual combinations with minus signs between the first and
second terms in parentheses. Given
$\gamma^{12}\epsilon_{(2)}=\gamma^{56}\epsilon_{(2)}=\epsilon_{(3)}$, 
$\gamma^{14}\epsilon_{(3)}=\gamma^{58}\epsilon_{(3)}=\epsilon_{(1)}$,
and
$\gamma^{13}\epsilon_{(1)}=\gamma^{57}\epsilon_{(1)}=\epsilon_{(2)}$,  
together with 
$\gamma^{12}\epsilon_{(1)}=-\gamma^{56}\epsilon_{(1)}=\epsilon_{(4)}$, 
$\gamma^{14}\epsilon_{(2)}=-\gamma^{58}\epsilon_{(2)}=\epsilon_{(4)}$,
and
$\gamma^{13}\epsilon_{(3)}=-\gamma^{57}\epsilon_{(3)}=\epsilon_{(4)}$,  
we have the explicit expressions
\begin{equation}
\begin{aligned}
   J^1 &= e^{12} + e^{34} , & \qquad 
   J'{}^1 &= e^{56} + e^{78} , \\
   J^2 &= e^{14} + e^{23} , & \qquad
   J'{}^2 &= e^{58} + e^{67} , \\
   J^3 &= e^{13} + e^{42} , & \qquad
   J'{}^3 &= e^{57} + e^{86} .
\end{aligned} 
\end{equation}
Again, the corresponding volume form is given by~\eqref{d8vol} as
above. Note there are six $\SU(4)$-structures given by $J^A_\pm=J^A\pm
J'{}^A$ and similarly each spinor $\epsilon_{(a)}$ defines a
corresponding $\Spin(7)$-structure given by 
\begin{equation}
\begin{aligned}
   \Psi^{(1)} &= \vol + \vol' - J^1 \wedge J'{}^1 + J^2 \wedge J'{}^2
      + J^3 \wedge J'{}^3 , \\
   \Psi^{(2)} &= \vol + \vol' + J^1 \wedge J'{}^1 - J^2 \wedge J'{}^2
      + J^3 \wedge J'{}^3 , \\
   \Psi^{(3)} &= \vol + \vol' + J^1 \wedge J'{}^1 + J^2 \wedge J'{}^2
      - J^3 \wedge J'{}^3 , \\
   \Psi^{(4)} &= \vol + \vol' - J^1 \wedge J'{}^1 - J^2 \wedge J'{}^2
      - J^3 \wedge J'{}^3 .
\end{aligned}
\end{equation}

\smallsec{$G_2$} A $G_2$-structure defines a single invariant spinor
in $d=7$. This can be defined by the projections
\begin{equation}
\label{G2projs}
   \gamma^{1234}\epsilon = \gamma^{1256}\epsilon 
     = \gamma^{1357}\epsilon = - \epsilon ,
\end{equation}
where we have taken $\ii\gamma_{1\cdots7}=\id$. The associative
three-form~\eqref{g2form} is then given by  
\begin{equation}
   \phi_{mnp} = - \ii \bar{\epsilon} \gamma_{mnp} \epsilon .
\end{equation}
The corresponding volume form is given by 
\begin{equation}
   \vol_{m_1\dots m_7} = 
      \ii \bar{\epsilon} \gamma_{m_1\dots m_7} \epsilon .
\end{equation}
Note the relation between $\phi$ and $\vol$ is slightly
non-standard. It is the opposite to the conventions given, for
instance in~\cite{Joyce}. To match the expressions in~\cite{Joyce},
one replaces $e_7$ with $-e_7$ and permutes the new basis
$\vol=-e_{1234567}$ to $e_{3254761}$. Note that one can choose an
imaginary basis for the $\gamma$-matrices where
$\bar{\epsilon}=\epsilon^\trsp$. 

Lifting to $d=8$, the $G_2$-structure defines a pair of real spinors
$\epsilon_{(a)}$ with $a=1,2$ satisfying~\eqref{G2projs} of opposite
chirality. They can be distinguished by
\begin{equation}
   \gamma^{5678}\epsilon_{(1)}=-\epsilon_{(1)} , \qquad
   \gamma^{5678}\epsilon_{(2)}=+\epsilon_{(2)} .
\end{equation}
The $G_2$-structure is defined by $\phi$ and $K$ given by 
\begin{equation}
\begin{aligned}
   \phi_{mnp} &= - \bar{\epsilon}_{(1)}\gamma_{mnp}\epsilon_{(2)} , \\
   K_m &= \bar{\epsilon}_{(1)}\gamma_m\epsilon_{(2)} .
\end{aligned}
\end{equation}
With $\gamma^8\epsilon_{(1)}=\epsilon_{(2)}$, we have $K=e^8$ and
$\phi$ takes the standard form~\eqref{g2form}. The corresponding
volume form $\vol=e^1\wedge\dots\wedge e^8$ is given by 
\begin{equation}
   \vol_{m_1\dots m_8} 
      = \bar{\epsilon}_{(1)}\gamma_{m_1\dots m_8}\epsilon_{(1)}
      = - \bar{\epsilon}_{(2)}\gamma_{m_1\dots m_8}\epsilon_{(2)}
\end{equation}
The two $\Spin(7)$-structures defined by $\epsilon_{(a)}$ are given by 
\begin{equation}
\begin{aligned}
   \Psi^{(1)} &= - i_K {*\phi} + \phi \wedge K , \\
   \Psi^{(2)} &= - i_K {*\phi} - \phi \wedge K .
\end{aligned}
\end{equation}
Note with these conventions, $i_K{*\phi}=-{*_7\phi}$ where $*_7\phi$ is
the usual coassociative four-form, that is the Hodge dual of $\phi$ on
the seven-dimensional subspace orthogonal to $K$.

\smallsec{$\SU(3)$} The $\SU(3)$-structure defines a single chiral
complex spinor $\epsilon$. This can be defined by the conditions
\begin{equation}
\label{SU3projs}
   \gamma^{1234}\epsilon = \gamma^{1256}\epsilon = - \epsilon .
\end{equation}
We choose the chirality $\ii\gamma^{1\dots6}\epsilon=\epsilon$ so
that $\gamma^{12}\epsilon=\ii\epsilon$. The forms $J$ and $\Omega$ are
then given by    
\begin{equation}
\begin{aligned}
   J_{mn} &= - \ii\bar{\epsilon}\gamma_{mn}\epsilon , \\
   \Omega_{mnp} &= \bar{\epsilon}^\conj\gamma_{mnp}\epsilon .
\end{aligned}
\end{equation}
Given $\gamma^{135}\epsilon=\epsilon^\conj$, we get the standard
expressions 
\begin{equation}
\label{JOform}
\begin{aligned}
   J &= e^{12} + e^{34} + e^{56} , \\
   \Omega &= (e^1+\ii e^2)(e^3+\ii e^4)(e^5+\ii e^6) . 
\end{aligned}
\end{equation}
The corresponding volume form is 
\begin{equation}
   \vol_{m_1\dots m_6} = \ii \bar{\epsilon} \gamma_{m_1\dots m_6} \epsilon .
\end{equation}
Again one can always choose a basis where
$\bar{\epsilon}=\epsilon^\dag$ and $\epsilon^\conj=\epsilon^*$.  

Lifting to $d=7$, the $\SU(3)$-structure defines a pair of invariant
spinors $\epsilon_{(a)}$ with $a=1,2$
satisfying~\eqref{SU3projs}. Fixing $\ii\gamma_{1\cdots7}=\id$, they
can be distinguished by 
\begin{equation}
   \gamma^{1357}\epsilon_{(1)}=-\epsilon_{(1)} , \qquad
   \gamma^{1357}\epsilon_{(2)}=+\epsilon_{(2)} .
\end{equation}
The $\SU(3)$-structure is given by 
\begin{equation}
\begin{aligned}
   J_{mn} &= - \bar{\epsilon}_{(1)}\gamma_{mn}\epsilon_{(2)} , \\
   \Omega_{mnp} &= 
      \ii \bar{\epsilon}_{(1)}\gamma_{mnp}\epsilon_{(2)} 
      - \tfrac{1}{2} \left(
         \bar{\epsilon}_{(1)}\gamma_{mnp}\epsilon_{(1)}
         - \bar{\epsilon}_{(2)}\gamma_{mnp}\epsilon_{(2)} 
      \right) , \\
   K_m &= - \ii \bar{\epsilon}_{(1)}\gamma_m\epsilon_{(2)} . 
\end{aligned}
\end{equation}
Given $\gamma^{12}\epsilon_{(1)}=\epsilon_{(2)}$, this gives $K=e^7$
and $J$ and $\Omega$ take the standard form~\eqref{JOform}. The corresponding
volume form $\vol=e^1\wedge\dots\wedge e^7$ is given by 
\begin{equation}
   \vol_{m_1\dots m_7} 
      = \ii \bar{\epsilon}_{(1)}\gamma_{m_1\dots m_7}\epsilon_{(1)}
      = \ii \bar{\epsilon}_{(2)}\gamma_{m_1\dots m_7}\epsilon_{(2)}
\end{equation}
The two $G_2$-structures defined by $\epsilon_{(a)}$ are given by 
\begin{equation}
\begin{aligned}
   \phi^{(1)} = J \wedge K - \im\Omega , \\
   \phi^{(2)} = J \wedge K + \im\Omega , \\
\end{aligned}
\end{equation}

\smallsec{$\SU(2)$} Finally for $\SU(2)$ the structure again defines a
single complex spinor of definite chirality. We take the negative
chirality 
\begin{equation}
   \gamma^{1234}\epsilon = - \epsilon . 
\end{equation}
The forms $J$ and $\Omega$ are then given by 
\begin{equation}
\begin{gathered}
   J_{mn} \equiv J^3_{mn} = - \ii\bar{\epsilon}\gamma_{mn}\epsilon , \\
   \Omega_{mn} \equiv J^2_{mn} + \ii J^1_{mn} 
      = \bar{\epsilon}^\conj\gamma_{mn}\epsilon .
\end{gathered}
\end{equation}
Given $\gamma^{12}\epsilon=\ii\epsilon$ and
$\gamma^{13}\epsilon=\epsilon^\conj$ we get the self-dual combinations
\begin{equation}
\label{JOform2}
\begin{aligned}
   J^1 &= e^{14} + e^{23} , \\
   J^2 &= e^{13} + e^{42} , \\ 
   J^3 &= e^{12} + e^{34} .
\end{aligned} 
\end{equation}
The corresponding volume form is 
\begin{equation}
   \vol_{m_1\dots m_4} = \ii \bar{\epsilon} \gamma_{m_1\dots m_4} \epsilon .
\end{equation}
Again one can always choose a basis where
$\bar{\epsilon}=\epsilon^\dag$ and $\epsilon^\conj=\epsilon^*$.

Lifting to $d=6$, the $\SU(2)$-structure defines a pair of complex
invariant spinors $\epsilon_{(a)}$ with $a=1,2$
satisfying~\eqref{SU3projs}. These have opposite chirality and can be
distinguished by  
\begin{equation}
   \gamma^{3456}\epsilon_{(1)}=-\epsilon_{(1)} , \qquad
   \gamma^{3456}\epsilon_{(2)}=+\epsilon_{(2)} .
\end{equation}
The $\SU(2)$-structure is given by 
\begin{equation}
\begin{gathered}
   J_{mn} = - \tfrac{1}{2} \ii \left(
         \bar{\epsilon}_{(1)}\gamma_{mn}\epsilon_{(1)}
         + \bar{\epsilon}_{(2)}\gamma_{mn}\epsilon_{(2)} 
      \right) , \\
   \Omega_{mn} = 
      \bar{\epsilon}_{(1)}^\conj\gamma_{mn}\epsilon_{(2)} , \\
   K^1_m + \ii K^2_m = \bar{\epsilon}_{(2)}\gamma_m\epsilon_{(1)} .
\end{gathered}
\end{equation}
Given $\gamma^{12}\epsilon_{(i)}=\epsilon_{(i)}$ and 
$\gamma^{135}\epsilon_{(i)}=\epsilon_{(i)}^\conj$ while  
$\gamma^5\epsilon_{(1)}=\epsilon_{(2)}$ and
$\gamma^6\epsilon_{(1)}=\ii\epsilon_{(2)}$, we have $K_1=e^5$,
$K_2=e^6$ and $J$ and $\Omega$ take the standard
form~\eqref{JOform2}. The corresponding volume form
$\vol=e^1\wedge\dots\wedge e^6$ is given by  
\begin{equation}
   \vol_{m_1\dots m_6} 
      = \ii \bar{\epsilon}_{(1)}\gamma_{m_1\dots m_6}\epsilon_{(1)}
      = - \ii \bar{\epsilon}_{(2)}\gamma_{m_1\dots m_6}\epsilon_{(2)}
\end{equation}
The two $\SU(3)$-structures defined by $\epsilon_{(a)}$ are given by 
\begin{equation}
\begin{aligned}
   J^{(1)} &= J + K^1 \wedge K^2 , &&& 
   \Omega^{(1)} &= \Omega \wedge (K^1 + \ii K^2) , \\
   J^{(2)} &= J - K^1 \wedge K^2 , &&&
   \Omega^{(2)} &= \Omega \wedge (K^1 - \ii K^2) . 
\end{aligned}
\end{equation}
%


\section{Almost product structures}
\label{app:product}


An almost complex structure is a $GL(n,\bbC)$-structure on a 
$2n$-dimensional 
manifold,
which is characterised by a tensor $J_m{}^n$ satisfying
$J\cdot J=-\id$. Using this one can split the tangent space $T_p M^\bbC$
at any point in the two subspaces $T_p M^+ \oplus T_p M^-$ corresponding 
to the $+\ii$ and $-\ii$ eigenvalues of $J$ respectively.
The Nijenhuis tensor for the almost complex structure is defined by
\bea
N_{mn}^{\;\;\;\;\;\;r} & = & J_m^{\;\;\;s} \, \de_{[s} J_{n]}^{\;\;\;r}-
J_n^{\;\;\;s}\, \de_{[s} J_{m]}^{\;\;\;r}~.
\eea
The almost complex structure is integrable if and only if
the Nijenhuis tensor vanishes and in this case one can 
introduce holomorphic co-ordinates on the manifold. 
If $J$ is compatible with a metric,
namely $J_{mq}\equiv J_m^{\;\;\;n} g_{nq}$ is a two-form, then the
metric is called almost Hermitian and Hermitian if $J$ is integrable.

Similarly, an almost product structure is a $GL(P,\bbR)\times
GL(Q,\bbR)$-structure on a $P+Q$-dimensional manifold, which is
characterised by a tensor $\Pi_m{}^n$ satisfying
$\Pi\cdot \Pi=+\id$. At any point the tangent space splits accordingly as
$T_p M= T_p M^P \oplus T_p M^Q$, where $P$ (respectively $Q$) is the 
number
of $+1$ (respectively $-1$) eigenvalues of $\Pi$.
The Nijenhuis tensor for the almost product structure is defined again by
\bea
N_{mn}^{\;\;\;\;\;\;r} & = & \Pi_m^{\;\;\;s} \, \de_{[s} 
\Pi_{n]}^{\;\;\;r}-
\Pi_n^{\;\;\;s}\, \de_{[s} \Pi_{m]}^{\;\;\;r} 
\eea
and the almost product structure is integrable if and only if
the Nijenhuis tensor vanishes (see e.g. \cite{yano}). If furthermore 
the almost product structure is metric compatible, i.e. 
$\Pi_{mq}\equiv\Pi_m^{\;\;\;n} g_{nq}$ is a symmetric tensor, 
one can introduce ``separating co-ordinates'' on the manifold such that the 
metric 
takes the $(P\times P, Q\times Q)$ block-diagonal form 
\bea
\dd s^2 & = & g^P_{ij}(x,y) \,\diff x^{i} \diff x^{j} 
+ g^Q_{ab}(x,y)\, \diff y^{a}\diff y^{b} 
\eea
where $i,j=1,\dots ,P$ and $a,b=1,\dots ,Q$.

Two commuting almost complex structures $J,J'$, satisfying
$J\cdot J'=J'\cdot J$ give rise to an almost product
structure
\bea\label{defalps}
\Pi=J\cdot J' .
\eea
Suppose $J$ and $J'$ are metric compatible and satisfy $\nabla^+
J=\nabla^+ J'=0$, or $\nabla^-J=\nabla^- J'=0$,
where $\nabla^\pm$ is a metric connection with totally
anti-symmetric torsion $\pm\frac{1}{2}H$. The Nijenhuis tensor then
reads in general   
\bea
N_{mn}^{\;\;\;\;\;\;r} = \pm\frac{1}{2} \left(H_{mn}^{\;\;\;\;\;\;r}+
\Pi_m{}^p \Pi_n{}^q H_{pq}^{\;\;\;\;\;r}  - \Pi^{rp}\Pi_m{}^q H_{pqn}-
\Pi_n{}^p\Pi^{rq} H_{pqm}\right) ~.
\eea
Using the tangent space decomposition, one finds that the only non-zero 
components are given by
\bea
N_{ij}^{\;\;\;\; c} &=& \pm 2\, H_{ij}^{\;\;\;\; c}\nn
N_{ab}^{\;\;\;\; k} &=& \pm 2\, H_{ab}^{\;\;\;\; k} . 
\label{nonzeronij}
\eea

If instead we assume that $J^+$, $J^-$ are commuting and are both integrable,
and also $\nabla^+ J^+=\nabla^- J^-=0$, then all the components of 
$N_{mn}{}^{r}$ vanish, hence $\Pi$ is integrable~\cite{ghr}. To see
this we first note that given the assumptions, $H$ is a $(2,1)+(1,2)$
form with respect to either complex structure $J^\pm$:
\begin{equation}
\label{twoone}
H_{mnr} = J_m{}^p J_n{}^q H_{pqr} + J_r{}^p J_m{}^q H_{pqn} + J_n{}^p 
J_r{}^q
H_{pqm} .
\end{equation}
To proceed, write $2\Pi=J^+\cdot J^- +J^-\cdot J^+$ to get
\bea
2\nabla_m \Pi_n{}^p=
J^+_n{}^r J^{-sp}H_{mrs}
-J^-_n{}^r J^{+sp}H_{mrs}
\eea
Then using \p{twoone} we find
\bea
4\nabla_{[m}\Pi_{n]}{}^p=
-J^+_m{}^rJ^+_n{}^sH_{rst}\Pi^{tp}
+J^-_m{}^rJ^-_n{}^sH_{rst}\Pi^{tp}
\eea
from which it easily follows that $N(\Pi)=0$.

It is sometimes incorrectly stated in the literature 
(see for instance~\cite{ghr,Sevrin:1996jr,LZ}) that $\Pi$, defined by
\p{defalps}, is integrable if and only if the two commuting almost
complex structures are integrable. A concrete class of counter-example
is provided by the geometry~\p{n=3} for generic instantons $G$.
This geometry has an $\SU(2)$ structure, built from the $\e^-$ Killing
spinors, which can be specified by two $\SU(3)$ structures.
The corresponding two almost complex structures, written as two-forms, 
are given by
\begin{equation}
\begin{aligned}
   J &= \ex^{2\Phi}(\dd x^1\wedge \dd x^2-\dd x^3\wedge \dd x^4)
         + (\dd y+B^1)\wedge (\dd z+ B^2) , \\
   J' &= \ex^{2\Phi}(\dd x^1\wedge \dd x^2-\dd x^3\wedge \dd x^4)
         -(\dd y+B^1)\wedge (\dd z+ B^2) .
\end{aligned}
\label{JJ'}
\end{equation}
Both almost complex structures are integrable.  
A quick way to see this
is to note that the geometry is a special example of the canonical $SU(3)$
geometry in $d=6$  (preserving twice as much supersymmetry) 
that was  discussed in section~\ref{sec:canon} (with expressions for
$\e^+$ spinors rather than $\e^-$ spinors that we have here)
for either $SU(3)$ structure. In particular, as pointed out 
in section 3, the almost complex structures are integrable. 
Moreover, the two complex structures clearly commute and thus
define an almost product structure given by
$\Pi=J\cdot J'$. 
On the other hand, because $\nabla^-J=\nabla^-J'=0$ and hence
$\nabla^-\Pi=0$, from \reef{nonzeronij}
we see that there are non-zero components of the associated 
Nijenhuis tensor, namely
\begin{equation}
\begin{aligned}
N_{m n}^{\;\;\;\;\;\; y} &=  -2\,G^1_{mn} \\
N_{m n}^{\;\;\;\;\;\; z} &=  -2\,G^2_{mn} .
\end{aligned}\label{Nyz}
\end{equation}

For definiteness, let us briefly present a simple example very
explicitly. In particular, set the dilaton field to zero and
$B^1=B^2=x^1 \diff x^2 +x^3 \diff x^4$. Then the almost complex
structures corresponding to \reef{JJ'} read  
\begin{equation}
   J_a{}^b =
      \begin{bmatrix}
         0 & 1 & 0 & 0  & -\,x^1 & -\,x^1 \\
         -1 & 0 & 0 & 0  & -\, x^1 &  x^1\\
         0 & 0 & 0 & -1 &  x^3 & x^3\\
         0 & 0 & 1 & 0  & -\, x^3 &  x^3\\
         0 & 0 & 0 & 0 & 0 & 1\\
         0 & 0 & 0 & 0 & -1 & 0
      \end{bmatrix}
   \qquad
   J'_a{}^b = 
      \begin{bmatrix}
         0 & 1 & 0 & 0  & -\,x^1 & -\,x^1 \\
         -1 & 0 & 0 & 0  &  x^1 & -\, x^1\\
         0 & 0 & 0 & -1 &  x^3 & x^3\\
         0 & 0 & 1 & 0  &  x^3 & -\, x^3\\
         0 & 0 & 0 & 0 & 0 & 1\\
         0 & 0 & 0 & 0 & -1 & 0
      \end{bmatrix}
\end{equation}
It is not difficult to check directly that these are both integrable
and indeed commute. The corresponding almost product structure is
\begin{equation}
   \Pi =J \cdot J' =
      \begin{bmatrix}
         -1 & 0 & 0 & 0  & 0 & 0 \\
         0 & -1 & 0 & 0  & 2\,x^1 &  2\,x^1\\
         0 & 0 & -1 & 0 &  0 & 0 \\
         0 & 0 & 0 & -1 & 2\, x^3 &  2\, x^3\\
         0 & 0 & 0 & 0 & 1 & 0\\
         0 & 0 & 0 & 0 & 0 & 1
      \end{bmatrix}
\end{equation}
Computing the corresponding \nij\ tensor, we find that it has the
non-zero components given by~\p{Nyz} with $G^1=G^2=\dd x^1\wedge\dd
x^2 +\dd x^3\wedge\dd x^4$. It would be interesting to investigate the
consequences of this counter example, especially in the context of the
sigma model literature.



\end{document}